\documentclass[prd,twocolumn,altaffilletter,tightenlines,showpacs,showkeys,floatfix,
               notitlepage,preprintnumbers,nofootinbib,superscriptaddress]{revtex4-1}

\usepackage{color}
\usepackage[colorlinks=true,citecolor=purple,linkcolor=blue]{hyperref}
\usepackage[normalem]{ulem}
\usepackage{amsmath,amssymb,slashed}
\usepackage[range-units=single,range-phrase=--]{siunitx}
\usepackage{bbm}
\usepackage{epsfig}
\usepackage{grffile,graphicx}
\usepackage{url}
\usepackage{enumerate}
\usepackage{multirow}
\usepackage{placeins}
\usepackage[dvipsnames]{xcolor}
\usepackage{epstopdf}
\usepackage[utf8]{inputenc}
\usepackage{tikz}
\usetikzlibrary{trees}
\usetikzlibrary{decorations.pathmorphing}
\usetikzlibrary{decorations.markings}

\definecolor{jblue}  {RGB}{20,50,100}
\definecolor{npurple}  {RGB} {153, 51, 204}
\definecolor{wred}   {RGB}{217,0,56}
\definecolor{white}   {RGB}{255,255,255}

\definecolor{korange}   {RGB}{235, 80,  43}
\definecolor{korange2}   {RGB}{245, 100,  63}
\definecolor{kyelloworange}   {RGB}{255, 210,  110}
\definecolor{kyelloworange2}   {RGB}{240, 170,  90}
\definecolor{kred}   {RGB}{204,  102, 153}
\definecolor{kpurple}   {RGB}{153,  61, 190}
\definecolor{kpurplelight}   {RGB}{213,  161, 230}

\usepackage{cleveref}
\usepackage{subfigure} 
\usepackage{lipsum}
\definecolor{red}{rgb}{1.0, 0, 0}
 \usepackage{gensymb}
\allowdisplaybreaks

\newcommand{\bra}[1]{\ensuremath{\langle #1 |}}   
\newcommand{\ket}[1]{\ensuremath{| #1 \rangle}}   
\newcommand{\sprod}[2]{\ensuremath{\left\langle #1 |%
                     #2 \right\rangle}}  

\newcommand{\diag}{\text{diag}}


\begin{document}

\title{Decaying Sterile Neutrinos and the Short Baseline Oscillation Anomalies}
\author{Mona Dentler}    \email{monainge.dentler@uni-goettingen.de}
\affiliation{Institut f\"{u}r Astrophysik,
             Georg August-Universit\"{a}t G\"{o}ttingen, Germany}

\author{Ivan Esteban}    \email{ivan.esteban@fqa.ub.edu}
\affiliation{Departament de Fis\'{i}ca Qu\`{a}ntica i Astrof\'{i}sica and
             Institut de Ciencies del Cosmos, Universitat de Barcelona, Spain}

\author{Joachim Kopp}    \email{jkopp@cern.ch}
\affiliation{Theoretical Physics Department, CERN, Geneva, Switzerland}
\affiliation{PRISMA Cluster of Excellence and
             Mainz Institute for Theoretical Physics,
             Johannes Gutenberg-Universit\"{a}t Mainz, Germany}

\author{Pedro Machado}   \email{pmachado@fnal.gov}
\affiliation{Fermi National Accelerator Laboratory, Batavia, Illinois, USA}
\date{\today}

\preprint{FERMILAB-PUB-19-538-T}

\begin{abstract}
The MiniBooNE experiment has observed a significant excess of electron 
neutrinos in a muon neutrino beam, in conflict with standard neutrino 
oscillations. We discuss the possibility that this excess is explained 
by a sterile neutrino with a mass $\sim \SI{1}{keV}$ that decays quickly 
back into active neutrinos plus a new light boson. This scenario 
satisfies terrestrial and cosmological constraints because it has 
neutrino self-interactions built-in. Accommodating also the LSND, 
reactor, and gallium anomalies is possible, but requires an extension of 
the model to avoid cosmological limits.
\end{abstract}

\maketitle

\section{Introduction}
\label{sec:intro}

Many major discoveries in neutrino physics have started out as oddball
anomalies that gradually evolved into incontrovertible evidence.  In this
work, we entertain the possibility that history is repeating itself in the
context of the MiniBooNE anomaly.  From 2002 to 2019, the MiniBooNE experiment
has been searching for electron neutrinos ($\nu_e$) appearing in a muon
neutrino ($\nu_\mu$) beam~\cite{AguilarArevalo:2010wv, Aguilar-Arevalo:2013pmq,
Aguilar-Arevalo:2018gpe},\footnote{Here and in the following, when we
say neutrino we mean also the corresponding anti-neutrinos.} and has found a
corresponding signal at $4.8\sigma$ statistical significance.  For some time,
the simplest explanation for this signal appeared to be the existence of a
fourth neutrino species $\nu_s$, called ``sterile neutrino'' because it would
not couple to any of the Standard Model interactions, but would communicate
with the Standard Model only via neutrino mixing. If $\nu_s$ has small but
non-zero mixing with both $\nu_e$ and $\nu_\mu$ and if the corresponding mostly
sterile neutrino mass eigenstate $\nu_4$ is somewhat heavier ($\sim 1$~eV) than
the Standard Model neutrinos, the MiniBooNE signal could be explained. This
explanation would also be consistent with a similar $3.8\sigma$ anomaly from
the earlier LSND experiment \cite{Aguilar:2001ty}, and with several reported
hints for anomalous disappearance of electron neutrinos in reactor experiments
\cite{Mention:2011rk, Dentler:2017tkw} and in experiments using intense
radioactive sources \cite{Acero:2007su, Giunti:2010zu}.\footnote{The latter
class of experiments is usually referred to as ``gallium experiments'', based
on the active component of their target material.} However, the sterile
neutrino parameter space consistent with MiniBooNE and these other anomalies is
in severe tension with the non-observation of anomalous $\nu_\mu$ disappearance
\cite{Kopp:2011qd, Conrad:2012qt, Archidiacono:2013xxa, Kopp:2013vaa,
Mirizzi:2013kva, Giunti:2013aea, Gariazzo:2013gua, Collin:2016rao,
Gariazzo:2017fdh, Giunti:2017yid, Dentler:2018sju}, unless several additional
new physics effects are invoked concomitantly~\cite{Liao:2018mbg, Moulai:2019gpi}.

In this work, we propose a different explanation for the MiniBooNE anomaly,
and possibly also for the LSND, reactor, and gallium anomalies.
In particular, we consider a sterile neutrino that rapidly decays back into
Standard Model (``active'') neutrinos $\nu_a$ \cite{PalomaresRuiz:2005vf, Bai:2015ztj,deGouvea:2019qre}.
The MiniBooNE excess is then interpreted
as coming from these decay products.  We will see that this scenario requires
only very small mixing between $\nu_s$ and $\nu_\mu$, thus avoiding the strong
$\nu_\mu$ disappearance constraints.  It also requires somewhat larger mixing between
$\nu_s$ and $\nu_e$, in line with the hints from reactor and radioactive source
experiments.  Finally, we will argue that decaying sterile neutrinos may avoid
cosmological constraints because the model automatically endows sterile
neutrinos with self-interactions (``secret interactions'' \cite{Hannestad:2013ana,
Dasgupta:2013zpn}).

\section{Decaying Sterile Neutrino Formalism}
\label{sec:formalism}

We extend the Standard Model by a sterile neutrino $\nu_s$ (a Dirac fermion)
and a singlet scalar $\phi$. The relevant interaction and
mass terms in the Lagrangian of the model are
\begin{align}
  \mathcal{L} \supset - g \, \bar\nu_{s} \nu_{s} \phi
                      - \sum_{a=e,\mu,\tau,s} m_{\alpha\beta} \,
                        \bar\nu_{\alpha} \nu_{\beta} \,.
  \label{eq:L-fermion}
\end{align}
The neutrino flavor eigenstates $\nu_\alpha$ are linear combinations of the mass
eigenstates $\nu_j$ ($j=1,2,3,4$) according to the relation $\nu_\alpha =
U_{\alpha j} \nu_j$, where $U$ is the unitary $4 \times 4$ leptonic mixing
matrix.  The first term in \cref{eq:L-fermion} can thus be rewritten as
\begin{align}
  - g \, \bar\nu_F \nu_F \phi
  - g \, |U_{s4}|^2 \bar\nu_4 \nu_4 \phi
  - (g \, U_{s 4}^* \bar\nu_4 \nu_F \phi + h.c.) \,,
  \label{eq:int-Lagrangian}
\end{align}
with
\begin{align}
  \nu_F \equiv \sum_{i=1}^3 U_{si} \nu_i \,.
  \label{eq:nu-F}
\end{align}
We assume initially that the
fourth, mostly sterile, mass eigenstate $\nu_4 \simeq \nu_s$ has a mass $m_4$
between $\mathcal{O}(\si{eV})$ and $\mathcal{O}(\SI{100}{keV})$, and that the
mass of $\phi$ is of the same order, but smaller.  The last term in
\cref{eq:int-Lagrangian} will then induce $\nu_4 \to \nu_F + \phi$ decays,
while the first term is responsible for $\phi \to \nu_F + \bar{\nu}_F$ decays.
When these decays occur in a neutrino beam, they will produce lower-energy
neutrinos at the expense of higher-energy ones, and they may also alter the
flavor structure of the beam.  In particular, they can produce excess
low-energy $\nu_e$ in a $\nu_\mu$ beam, as suggested by the MiniBooNE
anomaly.

The phenomenology of the model depends mainly on five new parameters.  Besides
$m_4$ and $m_\phi$, these are the coupling $g$ and the mixings $|U_{e4}|^2$,
$|U_{\mu 4}|^2$ between $\nu_4$ and $\nu_e$, $\nu_\mu$.  
We will assume the mixing with $\nu_\tau$  to be zero and neglect the complex
phases, as these parameters do not play an important role in explaining
the MiniBooNE excess.
For practical purposes, it is convenient to quote $m_4
\Gamma_4$ instead of $g$, as $m_4 \Gamma_4$ appears directly in the laboratory
frame decay length $E / (m_4 \Gamma_4)$.  Also, it is convenient to use the
ratio $m_\phi / m_4$ instead of just $m_\phi$ because the ratio measures more
directly the kinematic suppression in $\nu_4$ decays.

The evolution in energy $E$ and time $t$ of a neutrino beam in our model can be
described by a neutrino density matrix $\hat\rho_\nu(E,x)$ (a $4 \times 4$ matrix
in flavor space), the corresponding antineutrino density matrix
$\bar{\hat\rho}_\nu(E,x)$, and the scalar density function $\rho_\phi(E, t)$.  The
evolution equations are~\cite{GonzalezGarcia:2005xw, Moss:2017pur},

\begin{align}
  \frac{d\hat\rho_\nu(E, t)}{dt}  &= - i [\hat{H}, \hat\rho_\nu]
                                 - \frac{1}{2} \big\{ \tfrac{m_4}{E} \hat\Gamma, \rho \big\}
                                 + \mathcal{R}_\nu[\hat\rho_\nu, \rho_\phi, E, t]
  \label{eq:eom-rho-nu} \\
  \frac{d\rho_\phi(E, t)}{dt} &= - \tfrac{m_\phi}{E} \Gamma_\phi \rho_\phi
                                 + \mathcal{R}_\phi[\hat\rho_\nu, E, t]
  \label{eq:eom-rho-phi}
\end{align}

\noindent
where $\hat{H} = \frac{1}{2E} \, \diag(0, \Delta m_{21}^2, \Delta m_{31}^2,
\Delta m_{41}^2)$ is the standard neutrino oscillation Hamiltonian, written
here in the mass basis, and $\hat\Gamma = \Gamma_4 \hat\Pi_4$ is the decay
term, which contains the projection operator $\hat\Pi_4 =
\ket{\nu_4}\bra{\nu_4}$ onto the fourth, mostly sterile, mass eigenstate as
well as the decay width $\Gamma_4$ of $\nu_4$ in its rest frame.  Similarly,
$\Gamma_\phi$ is the rest frame decay width of $\phi$.  The functional
$\mathcal{R}_\nu[\hat\rho_\nu, \rho_\phi, E, t]$ describes the appearance of
the daughter neutrinos from $\nu_4$ and $\phi$ decay.  Neglecting the masses of
$\nu_1$, $\nu_2$, and $\nu_3$, it is given by
\begin{widetext}
\begin{align}
  \mathcal{R}_\nu[\hat\rho_\nu, \rho_\phi, E, t]
    &= \hat{\Pi}_F \int_{\frac{E}{1-x_{\phi_4}^2}}^{\infty} \! dE_4 \sum_k
          \hat\rho_{\nu,44}(E_4, t) \, \frac{d\Gamma^\text{lab}(\nu_4 \to \nu_k \phi)}{dE_k}
    \;+\; \hat{\Pi}_F \sum_{k, j} \int_E^{\infty} \! dE_\phi \,
          \rho_\phi(E_\phi, t) \frac{d\Gamma^\text{lab}(\phi \to \nu_k \bar{\nu}_j)}{dE_k} \,,
  \label{eq:R-nu}
\end{align}
where $d\Gamma^\text{lab}(X \to Y)/dE_k$ are the differential decay widths for
the various decays $X \to Y$ in the lab frame, and $x_{\phi 4} \equiv
m_\phi/m_4$. The projection operator 
\begin{align}
  \hat\Pi_F = \frac{\ket{\nu_F} \bra{\nu_F}}{|\!\sprod{\nu_F}{\nu_F}\!|^2}
            = \sum_{i,j=1}^3 \frac{U_{si}^* U_{sj}}{\sum_k |U_{sk}|^2} \ket{\nu_i}\bra{\nu_j}
  \label{eq:Pi-F}
\end{align}
isolates the specific combination of mass eigenstates that appears in $\nu_4$ and $\phi$
decays, and the integrals run over all
parent energies $E_4$, $E_\phi$ that lead to daughter neutrinos of energy
$E$. Analogously, $\mathcal{R}_\phi[\hat\rho_\nu, E, t]$ describes the appearance
of scalars from $\nu_4$ decay:
\begin{align}
  \mathcal{R}_\phi[\hat\rho_\nu, E, t]
    &= \int_{E}^{E/x_{\phi 4}^2} \! dE_4 \sum_k \bigg[
         \hat\rho_{\nu,44}(E_4, t) \, \frac{d\Gamma^\text{lab}(\nu_4 \to \nu_k \phi)}{dE_\phi}
       + \bar{\hat\rho}_{\nu,44}(E_4, t) \frac{d\Gamma^\text{lab}(\bar{\nu}_4 \to \bar\nu_k \phi)}{dE_\phi}
       \bigg] \,.
  \label{eq:Rphi}
\end{align}
\end{widetext}
\noindent
With the appearance terms $\mathcal{R}_\nu[\hat\rho_\nu, \rho_\phi, E, t]$ and
$\mathcal{R}_\phi[\hat\rho_\nu, E, t]$ defined, the equations of motion
\eqref{eq:eom-rho-nu} and \eqref{eq:eom-rho-phi} can be solved analytically if we neglect matter effects.
Neglecting furthermore the small mass splittings between the three light
neutrino mass eigenstates, the electron neutrino flux $\phi_e(L, E)$ appearing
in a muon neutrino beam of energy $E$ after a distance $L$ due to oscillations
and decay is given by
\begin{widetext}
\begin{align}
  \phi_e(L, E) = \phi_\mu(0,E) \, |U_{e4}|^2 |U_{\mu 4}|^2
                 \bigg[1 + e^{-\frac{m_4 \Gamma_4 L}{E}}
                         - 2 e^{-\frac{m_4 \Gamma_4 L}{2E}}
                                   \cos\bigg(\frac{\Delta m_{41}^2 L}{2E}\bigg) \bigg] 
                + |U_{\mu 4}|^2 \frac{|\sprod{\nu_e}{\nu_F}|^2}{|\sprod{\nu_F}{\nu_F}|^2}
                                                   \mathcal{I} \,.
  \label{eq:eom_solution}
\end{align}
Here, $\ket{\nu_F}$ is the superposition of mass eigenstates into which the
$\nu_4$ decay (defined in \cref{eq:nu-F}), and the decay integral $\mathcal{I}$
is given by
\begin{align}
  \begin{split}
    \mathcal{I}
      &= \int_{E/(1-x_{\phi4}^2)}^\infty \! dE_4 \,
         \Big(1 -  e^{-\frac{m_4 \Gamma_4 L}{E_4}} \Big) 
          \phi_\mu(0,E_4) \sum_j
          \frac{1}{\frac{m_4}{E_4}\Gamma_4}\frac{d\Gamma^\text{lab}(\nu_4 \to \nu_j \phi)}{dE}
                      \\
      &+ \int_E^\infty\!dE_\phi \int_{E_\phi}^{E_\phi/x_{\phi 4}^2} \! dE_4
         \frac{1}{\frac{m_4 \Gamma_4 L}{E_4} - \frac{m_\phi \Gamma_\phi L}{E_\phi}}
         \bigg[\Big(1 - e^{-\frac{m_\phi \Gamma_\phi L}{E_\phi}} \Big)
                 \frac{m_4 \Gamma_4 L}{E_4}
             - \Big(1 - e^{-\frac{m_4 \Gamma_4 L}{E_4}} \Big)
                 \frac{m_\phi \Gamma_\phi L}{E_\phi}
         \bigg] \\
      &\hspace{3cm} \times \frac{1}{\frac{m_4}{E_4} \Gamma_4}\sum_j
         \bigg[
           \phi_\mu(0,E_4) \frac{d\Gamma^\text{lab}(\nu_4 \to \nu_j \phi)}{dE}
         + \bar{\phi}_\mu(0,E_4) \frac{d\Gamma^\text{lab}(\bar{\nu}_4 \to \bar{\nu}_j \phi)}{dE}
         \bigg]  
         \sum_{i,j} \frac{1}{\frac{m_\phi}{E_\phi}\Gamma_\phi} \frac{d\Gamma^\text{lab}(\phi \to \nu_i \bar{\nu}_j)}{dE} \,.
  \end{split} 
  \label{eq:decay-integral}
\end{align}
\end{widetext}
\noindent
In the above equations, $\phi_\mu(0,E)$ and $\bar{\phi}_\mu(0, E)$ are the
initial $\nu_\mu$ and $\bar{\nu}_\mu$ fluxes, respectively. A completely
analogous equation describes $\bar{\nu}_e$ appearance.
 
The physical interpretation of \cref{eq:eom_solution} is straightforward: the
first term on the right-hand side describes $\nu_\mu \to \nu_e$ oscillations,
altered by the removal of neutrinos at energy $E$ due to $\nu_4$ decay.  In
fact, this contribution matches the result of
ref.~\cite{GonzalezGarcia:2008ru}, on invisible $\nu_4$ decay.  The second
term gives the contribution from neutrinos generated in $\nu_4$ and $\phi$
decays. The factor $|U_{\mu 4}|^2$ arises because $\nu_4$ is the only mass
eigenstate that decays. It describes the amount of $\nu_4$ in the $\nu_\mu$
beam. The factor $|\sprod{\nu_e}{\nu_F}|^2 / |\sprod{\nu_F}{\nu_F}|^2$ is the
probability of the decay product to be detected as an electron neutrino, and
the integral $\mathcal{I}$ controls the energy distribution of the decay
products.
 
Analytic expressions for the decay widths appearing in
\cref{eq:eom-rho-nu,eq:eom-rho-phi,eq:R-nu,eq:Pi-F,eq:Rphi,eq:eom_solution,eq:decay-integral}
are given in \cref{sec:decay-widths}.

\section{Fit to MiniBooNE data}
\label{sec:mb-fit}

To compare the predictions of the decaying sterile neutrino scenario to
MiniBooNE data, we evolve the unoscillated beam following the formulas given above.
We then follow the fitting procedure recommended by the MiniBooNE
collaboration (see the data releases accompanying
refs.~\cite{AguilarArevalo:2010wv, Aguilar-Arevalo:2018gpe}), but go beyond it
by accounting for the impact of $\nu_\mu$ and $\nu_e$ disappearance on the
signal and background normalization (see Appendix for details).

\begin{figure*}
  \centering
  \begin{tabular}{c@{\quad}c}
    \includegraphics[width=0.48\textwidth]{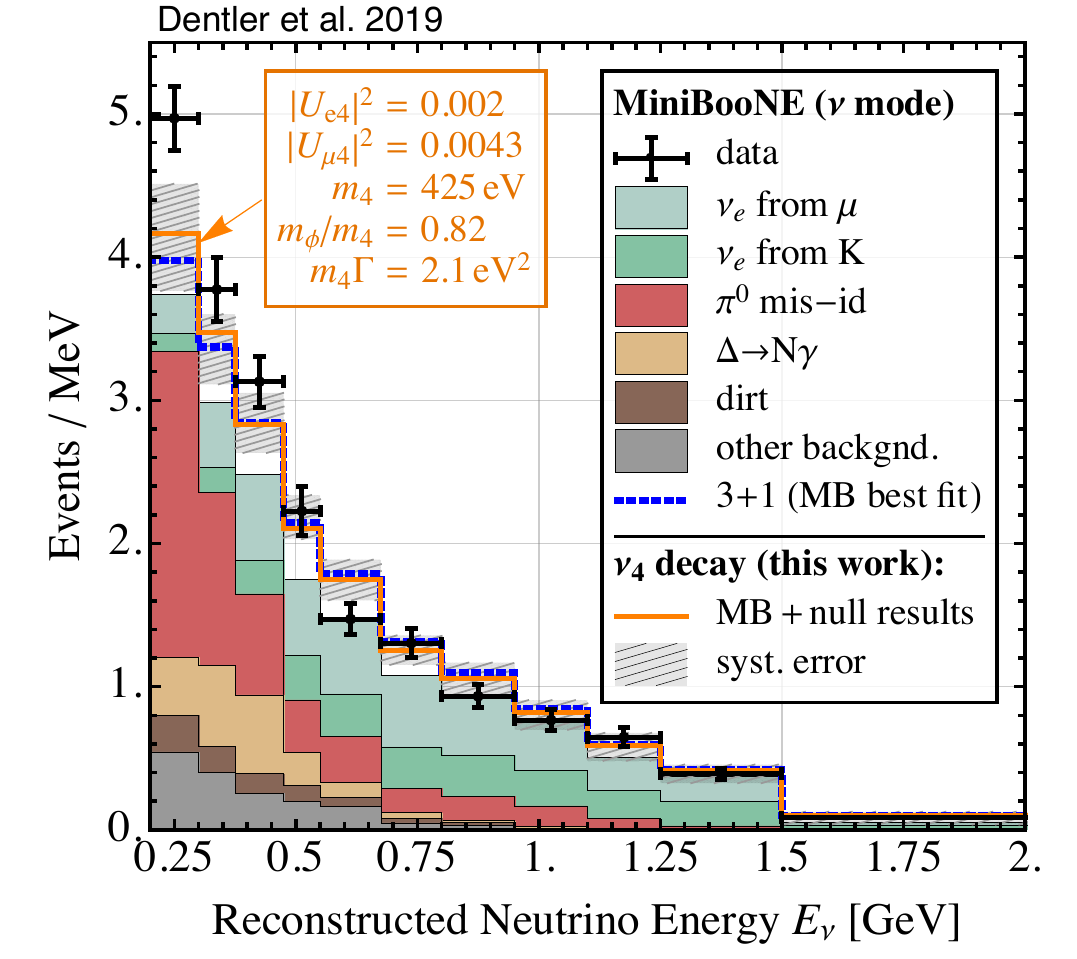} &
    \includegraphics[width=0.48\textwidth]{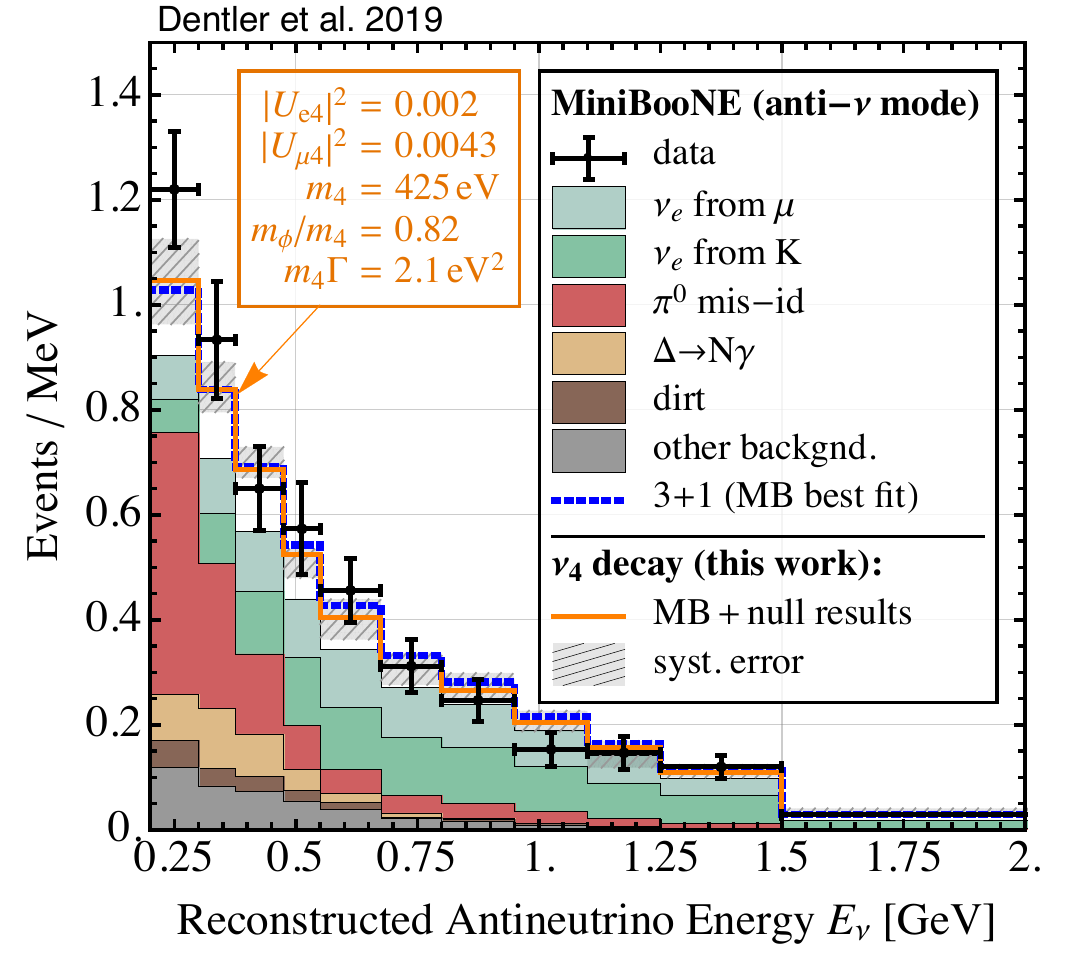}
  \end{tabular}
  \caption{Comparison of MiniBooNE neutrino-mode \emph{(left)} and
    anti-neutrino-mode \emph{(right)} data \cite{Aguilar-Arevalo:2018gpe}
    to the predictions of the neutrino oscillation + decay scenario discussed
    in this work.  We show the expected spectrum at the point which optimally
    fits MiniBooNE data, while being consistent with all null results
    (orange histogram with systematic error band; parameters given in the plot).
    We also show the MiniBooNE-only best point for $3+1$
    oscillations without decay (blue dotted histogram, parameter values
    $\Delta m_{41}^2 = \SI{0.13}{eV^2}$, $|U_{e4}|^2 = 0.024$,
    $|U_{\mu 4}|^2 = 0.63$).
  }
  \label{fig:mb-spectrum}
\end{figure*}

Illustrative results are shown in \cref{fig:mb-spectrum}, where we have chosen
parameter values that give an optimal fit to MiniBooNE data while being
consistent with null results from other oscillation experiments, as well as
non-oscillation constraints.  At $m_4 \Gamma_4 = \SI{2.1}{eV^2}$, most
$\nu_4$ will have decayed before reaching the detector.  The value $m_\phi /
m_4 = 0.82$ implies mild phase space suppression in $\nu_4$ decays, which tends
to shift the $\nu_e$ spectrum to lower energies, in excellent agreement with
the data. Compared to models with massless $\phi$ \cite{PalomaresRuiz:2005vf,
Bai:2015ztj}, our scenario also has the advantage that it allows $\phi
\to \nu_F \bar\nu_F$ decays, further boosting the $\nu_e$ flux at low energies.
It is therefore favored compared to the $m_\phi = 0$ case at more than
99\% confidence level.
The fit in our model is better than in oscillation-only
scenarios (blue dotted histogram in \cref{fig:mb-spectrum})
\cite{Dentler:2018sju}, which by themselves already offer an excellent fit
as long as only MiniBooNE data are considered (MiniBooNE quotes a
$\chi^2$ per degree of freedom of 9.9/6.7 \cite{Aguilar-Arevalo:2018gpe}).
Our model, however, is also consistent with all constraints.
Notably, it reproduces the
angular distribution of the neutrino interaction products in MiniBooNE because
it predicts an actual flux of electron neutrinos instead of attempting to mimic
the signal with other particles \cite{Gninenko:2009ks, Gninenko:2010pr,
Bertuzzo:2018itn, Ballett:2018ynz, Jordan:2018qiy, Fischer:2019fbw}.
In particular, the angle between the parent $\nu_s$ and the daughter $\nu_e$
is suppressed by a large Lorentz boost $\gamma \sim
\mathcal{O}(1\,000)$~\cite{Lindner:2001fx}.  This boost is sufficient
to ensure that the daughter neutrinos enter the MiniBooNE detector,
which is a $\sim \SI{6}{m}$ sphere located $\sim \SI{500}{m}$ from the
primary target, under essentially the same angle as the parent neutrino
would have done.

\section{Constraints}
\label{sec:constraints}

\begin{figure}
  \centering
  \includegraphics[width=1.0\columnwidth]{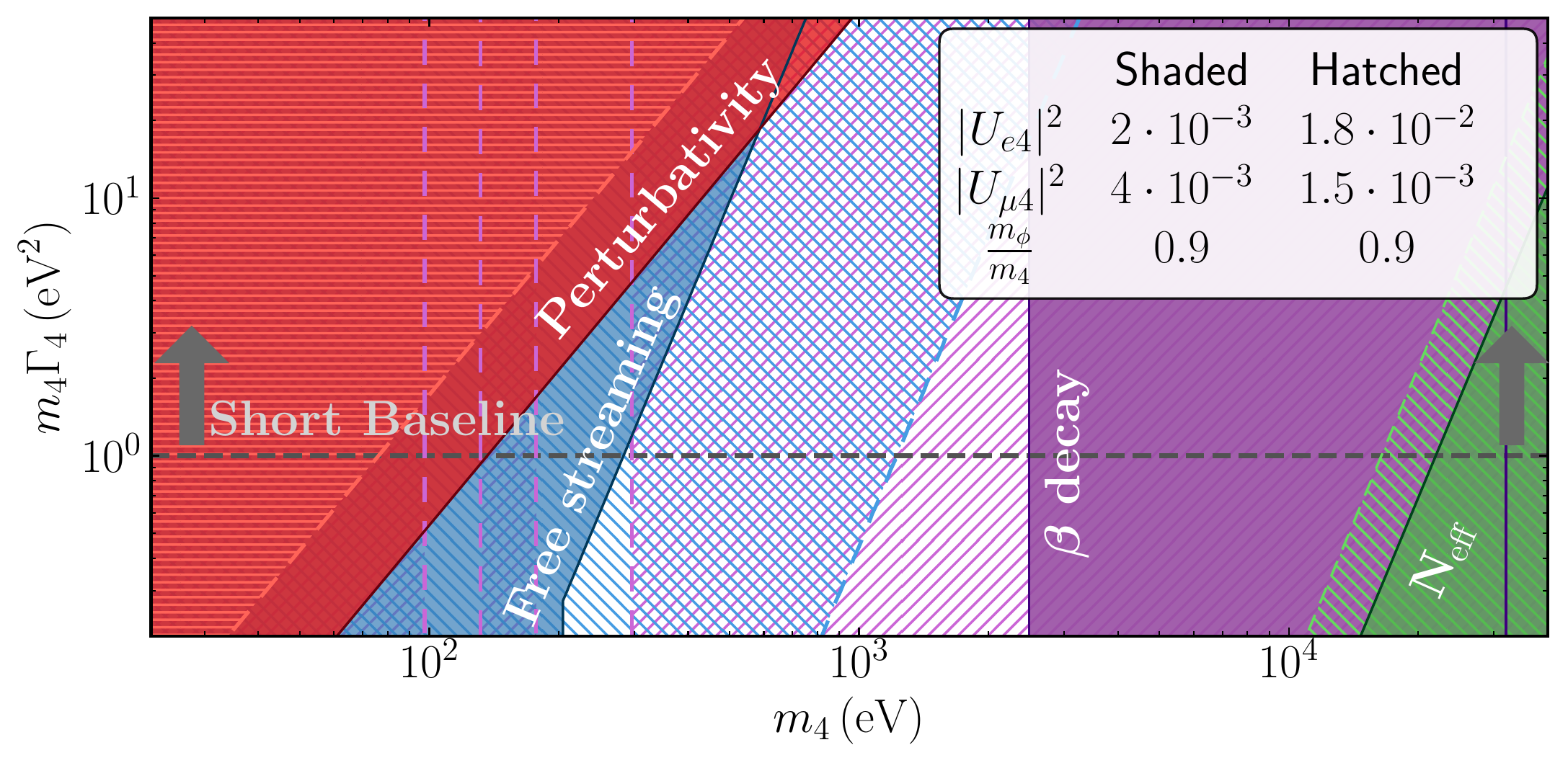}
  \caption{Non-oscillation constraints on decaying sterile neutrinos for
    parameters favored by the global fit without LSND (shaded), and by the
    global fit without the free-streaming constraint (hatched).}
  \label{fig:lab-constraints}
\end{figure}

\begin{figure*}
  \centering
  \begin{tabular}{c@{\quad}c}
    \includegraphics[width=0.48\textwidth]{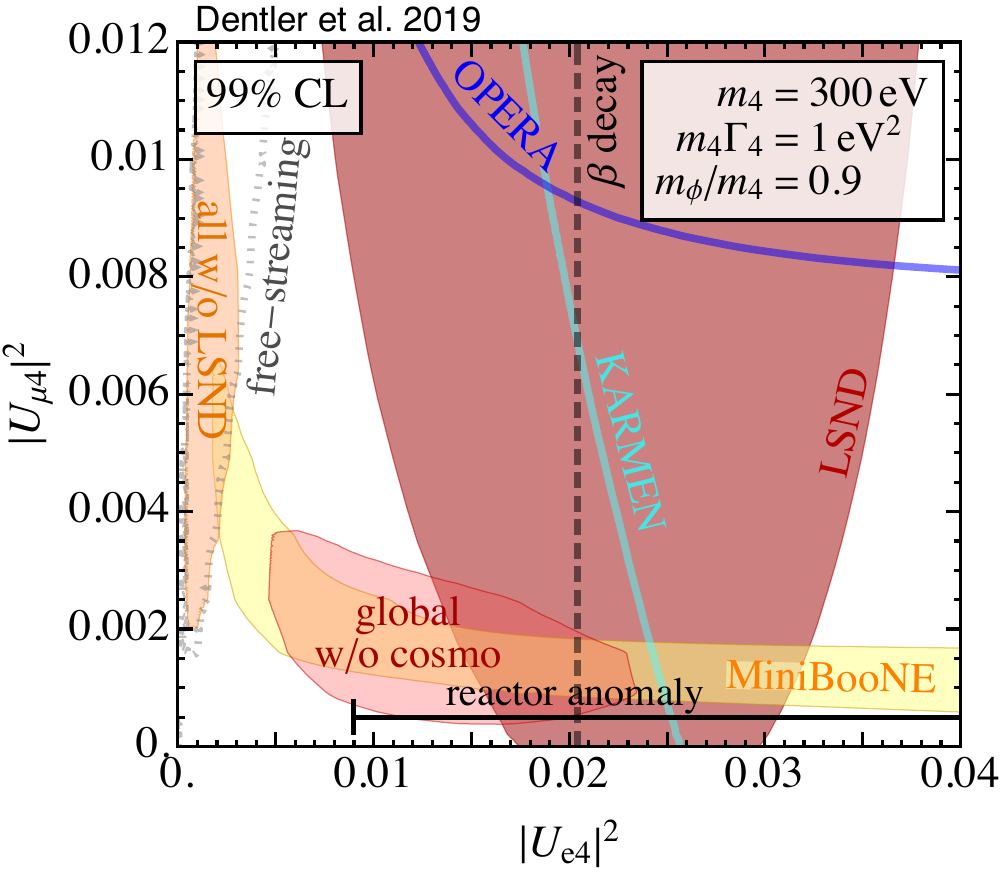} &
    \includegraphics[width=0.48\textwidth]{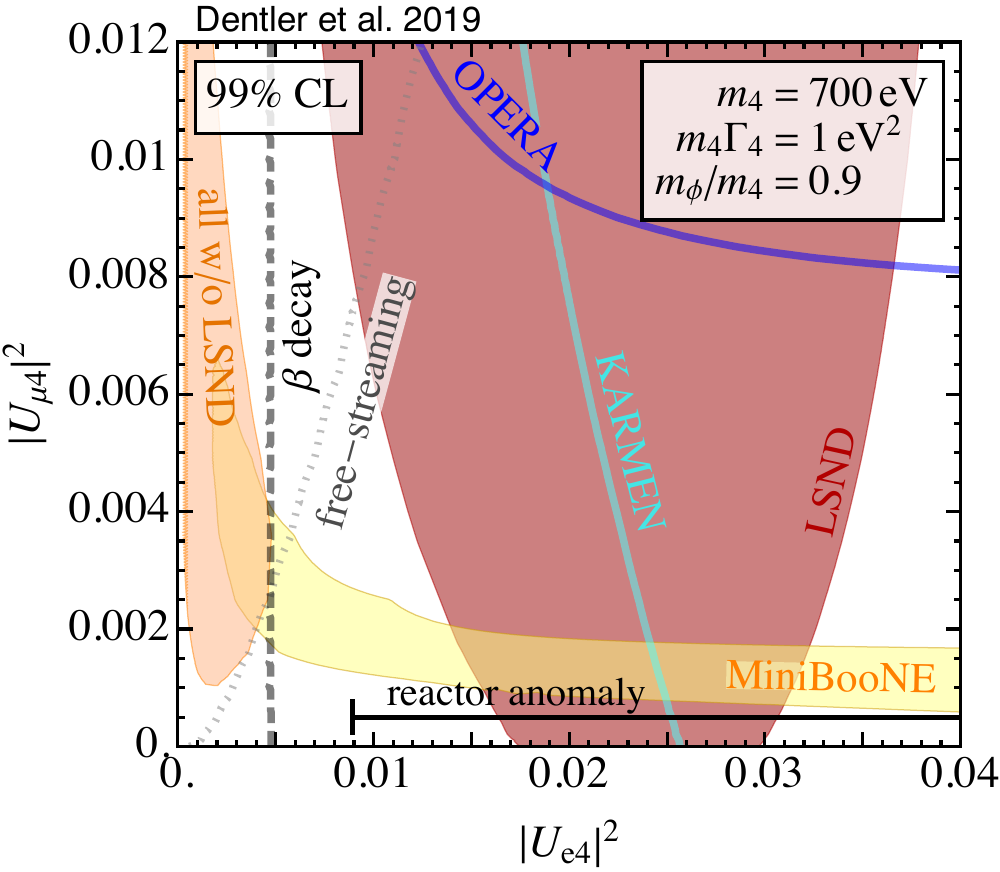}
  \end{tabular}
  \caption{Allowed values of the squared mixing matrix elements
    $|U_{e4}|^2$ and $|U_{\mu 4}|^2$ (measuring the mixing of $\nu_s$ with
    $\nu_e$ and $\nu_\mu$, respectively) in the decaying sterile neutrino
    scenario. We show two representative slices through the 5-dimensional 99\%
    confidence regions.  Our fits include MiniBooNE, OPERA, ICARUS,
    E776, and KARMEN data, as well as constraints from nuclear beta decay spectra
    and from the requirement of neutrino free-streaming in the early Universe.
    For the null results from oscillation experiments, the region to the right
    of the curves is excluded. For the free-streaming constraint, the region to
    the left of the gray dotted contour is excluded.
    We also show, as a black rule at the bottom of the plot, the $|U_{e4}|^2$
    range preferred by the reactor neutrino anomaly.    Constraints on
    $\nu_\mu$ disappearance are significantly weaker here than in the $3+1$
    scenario without decay, and are hence not shown.  We also do not show a fit
    including both LSND and cosmology as the goodness of fit would be very
    poor.  Note that the global combinations are sensitive to five degrees of
    freedom, namely $m_4$, $|U_{e4}|^2$, $U_{\mu 4}^2$, $m_4 \Gamma_4$, and
    $m_\phi / m_4$; oscillation experiments are sensitive only to the last four
    of these; beta decay spectra depend on two degrees of
    freedom ($m_4$ and $|U_{e4}|^2$); reactor experiments depend only on $|U_{e4}|^2$;
    and the free-streaming constraint depends only on the parameter combination
    $m_4 / |U_{s1}|$.}
  \label{fig:Ue4Um4}
\end{figure*}

We now discuss the various constraints that an explanation of the MiniBooNE
anomaly in terms of decaying sterile neutrinos has to respect.  The most
relevant constraints are also summarized in \cref{fig:Ue4Um4,fig:lab-constraints}.

{\it (1) Oscillation null results.}
Putting MiniBooNE into context with other $\nu_e$ appearance searches, we show
in \cref{fig:Ue4Um4} two slices through the 5-dimensional parameter space of
the decaying sterile neutrino model along the plane spanned by $|U_{e4}|^2$ and
$|U_{\mu 4}|^2$. To produce this figure, we have used fitting codes from
refs.~\cite{Kopp:2011qd, Kopp:2013vaa, Dentler:2018sju} (based partly on
refs.~\cite{Huber:2004ka, Kopp:2006wp, Huber:2007ji}). We see that most of the
parameter region preferred by MiniBooNE is well compatible with the KARMEN
short-baseline oscillation search \cite{Armbruster:2002mp} and with the OPERA
long-baseline experiment~\cite{Agafonova:2013xsk}.  We have checked that the
limits from ICARUS~\cite{Antonello:2012fu, Farese:2014, Antonello:2015jxa} and
E776~\cite{Borodovsky:1992pn} are significantly weaker.

All constraints on $|U_{e4}|^2$ ($|U_{\mu4}|^2$) from $\nu_e$ ($\nu_\mu$) disappearance 
experiments are avoided \cite{Gariazzo:2017fdh,Dentler:2018sju, Diaz:2019fwt}. This is 
mostly because in pure oscillation scenarios the number of excess events in MiniBooNE and 
LSND is proportional to $|U_{e4}|^2 |U_{\mu 4}|^2$, while in our scenario it is
proportional only to $|U_{\mu 4}|^2$ as long as $|U_{e4}|^2 \gg |U_{\mu 4}|^2$.
Therefore, it agrees well even with the tightest constraints~\cite{Adamson:2017uda,Louis:2018yeg}

We can already see from \cref{fig:Ue4Um4} that MiniBooNE is also compatible
with LSND and with the $|U_{e4}|^2$ range preferred by the reactor anomaly, but
only in a parameter region that would unacceptably reduce free-streaming of
active neutrinos in the early Universe. We will see below that this tension can
be avoided in extensions of the model.

{\it (2) Beta decay spectra} (purple regions in \cref{fig:lab-constraints} and
black dashed lines in \cref{fig:Ue4Um4}).
Direct searches for sterile neutrinos looking for anomalous features in beta
decay spectra \cite{Bryman:2019ssi, Atre:2009rg, Dragoun:2015oja,
deGouvea:2015euy} suggest that $\mathcal{O}(0.001-0.01)$ mixings
between active and sterile neutrinos -- as required by MiniBooNE -- are allowed
for $m_4 \lesssim \text{few keV}$.

{\it (3) Neutrinoless double beta decay.} If neutrinos are Majorana particles,
the non-observation so far of neutrinoless double beta decay requires
$m_4 |U_{e4}|^2 \lesssim \SI{0.2}{eV}$~\cite{Dolinski:2019nrj}. This is the
reason we always focus on Dirac neutrinos in this work.

{\it (4) $N_\text{eff}$}, a measure for the {\it energy density of relativistic
particles in the early Universe} (green region in \cref{fig:lab-constraints}).
The measured value of $N_\text{eff}$ is very
close to the SM value of $\sim 3$ both at the BBN and recombination epochs
\cite{Cyburt:2015mya, Aghanim:2018eyx}.
Naively, one might expect that this observation precludes the
existence of a fourth neutrino species with $m_4 \lesssim \si{MeV}$.
In our model, however, the $N_\text{eff}$
constraint is avoided by the ``secret interactions'' mechanism
\cite{Hannestad:2013ana, Dasgupta:2013zpn}: 
any small abundance of $\nu_s$ generates a temperature-dependent 
potential $V_\text{eff} \propto g^2 T$, reducing the $\nu_s$--$\nu_a$ 
mixing by a factor $\sqrt{\Delta m^2 / (E V_\text{eff})}$. Hence, 
the production of $\nu_s$ is suppressed until the
temperature drops low enough. For the parameter range that
the short-baseline anomalies are pointing to, this can easily be postponed to
late times ($T \ll\text{MeV}$), after neutrino-electron decoupling. Consequently, when $\nu_s$ are eventually produced, they are produced
at the expense of active neutrinos, so $N_\text{eff}$ does not change any
more and constraints are automatically satisfied.  More quantitatively,
$N_\text{eff}$ constraints are avoided when
\begin{align}
  (m_4 \Gamma_4)^\text{eff} \gtrsim \SI{2e-14}{eV^2} \bigg( \frac{m_4}{\si{eV}} \bigg)^4 \,,
  \label{eq:Neff-cons}
\end{align}
where we have defined
\begin{align}
  (m_4 \Gamma_4)^\text{eff} \equiv \frac{m_4 \Gamma_4}
                                        {|U_{s4}|^2
                                         (|U_{e4}|^2 + |U_{\mu 4}|^2)
                                         \Big( 1 - \tfrac{m_\phi^2}{m_4^2} \Big)^2} \,.
  \label{eq:m4G-eff}
\end{align}
This constraint can be easily satisfied in the mass range allowed by beta decay
limits.

{\it (5) $\sum m_\nu$, the sum of neutrino masses.} Massive
neutrinos affect the CMB as well as structure formation, and this has for
instance allowed the Planck collaboration to set a limit
$\sum m_\nu \lesssim \SI{0.12}{eV}$ \cite{Aghanim:2018eyx}.  In our
model, this constraint is easily satisfied because in the interesting
parameter range with $m_4 \gg \SI{1}{eV}$ and $m_4 \Gamma_4 \gtrsim \SI{1}{eV^2}$,
any $\nu_4$ that are produced in the early Universe will have
decayed via $\nu_4 \to \nu_{1,2,3} + (\phi \to \nu_{1,2,3}
\bar\nu_{1,2,3})$ long before recombination and the onset of structure formation.

{\it (6) Neutrino Free-Streaming} (blue region in \cref{fig:lab-constraints}
and gray dotted lines in \cref{fig:Ue4Um4}).  Via the mixing with $\nu_s$, also
the light neutrino mass eigenstates $\nu_{1,2,3}$ feel $\phi$-mediated
interactions and are therefore not fully free-streaming. This may put the model
in tension with CMB observations, which require that neutrinos should
free-stream from about redshift $10^5$ onwards \cite{Cyr-Racine:2013jua,
Forastieri:2017oma,Forastieri:2019cuf,Escudero:2019gfk,Escudero:2019gvw}.\footnote{It
  is noteworthy, though, that some cosmological
fits have actually found a \emph{preference} for neutrino self-interactions
\cite{Cyr-Racine:2013jua, Lancaster:2017ksf, Oldengott:2017fhy, Song:2018zyl,
Kreisch:2019yzn, Blinov:2019gcj} that could be accommodated in our model.}
This requirement bounds the squared coupling among the lightest neutrino 
mass eigenstate and the scalar $\phi$, i.e.,
$\left(g |U_{s1}|^2\right)^2$. (Heavier mass eigenstates are not relevant
as they decay quickly.)  Here we are taking $\nu_1$ to be the lightest
mass eigenstate, as favoured by current data. Quantitatively,
\begin{align}
  (m_4 \Gamma_4)^\text{eff} \lesssim
    \SI{4e-10}{eV^2} \bigg( \frac{m_4}{\text{eV}} \bigg)^4
                     \bigg( \frac{0.1}{|U_{s1}|} \bigg)^4 x_{\phi 4}^2 \,.
  \label{eq:free-streaming-constraint}
\end{align}
with $m_4 \lesssim \SI{200}{eV}$ required for $g^2 \gtrsim
10^{-6}$~\cite{Escudero:2019gvw}.  Note that in \cref{fig:Ue4Um4}, this
constraint is present even for very small mixings. This is because, at fixed
$m_4\Gamma_4$, small mixings need to be compensated by a large coupling $g$,
strengthening the free streaming constraint.  The value of $|U_{s1}|^2$ is
fixed in terms of $|U_{e1}|^2$ and $|U_{\mu 1}|^2$ by unitarity, assuming
the active neutrino mixing angles to be fixed at their values from
Ref.~\cite{Esteban:2018azc}.

However, the constraint could be substantially weakened in extensions of our
model, see for instance refs.~\cite{Bertuzzo:2018ftf, Zhao:2017wmo,
Chu:2018gxk, Farzan:2019yvo}. A minimalist example is the production of extra
species of light particles at the expense of the neutrino sector after neutrino
decoupling.  These would compensate for the lack of free-streaming in active
neutrinos.

{\it (7) SN 1987A.} The fact that neutrinos from supernova 1987A could be
observed at Earth without being absorbed through scattering on the cosmic
neutrino background constrains neutrino self-interactions \cite{Kolb:1987qy}. We have
checked that, due to mixing suppression, these constraints are avoided in our
scenario. Note that supernova cooling, which is sensitive to non-interacting sterile
neutrinos, does not constrain our model as $\nu_4$ and $\phi$ quickly
decay to lighter neutrinos that remain trapped in the supernova core.

{\it (8) Decays of SM neutrinos.}  We have checked that
decays of the form $\nu_{2,3} \to \bar\nu_1 + 2 \nu_1$, mediated
by an off-shell $\phi$, are always
sufficiently rare to be consistent with solar neutrino constraints
\cite{GonzalezGarcia:2008ru, Berryman:2014qha}.  Note, however,
that we predict the cosmic neutrino background today
to consist exclusively of $\nu_1$ or $\nu_3$, for normal and inverted neutrino mass ordering, respectively.

{\it (9) Perturbativity} (red region in \cref{fig:lab-constraints}).
Requiring that the $\nu_s$--$\phi$ coupling
constant $g$ in \cref{eq:L-fermion,eq:int-Lagrangian} is $< \sqrt{4\pi}$
imposes the bound
\begin{align}
  (m_4 \Gamma_4)^\text{eff} &\lesssim
    \SI{0.25}{eV^2} \bigg( \frac{m_4}{\si{eV}} \bigg)^2 \,.
\end{align}
Similarly to the free-streaming bound, this constraint applies even for
very small mixing when $m_4\Gamma_4$ is fixed.
This bound restricts $m_4$ in our model to be $\gtrsim \SI{100}{eV}$
for $m_4 \Gamma_4$ values large enough to explain the MiniBooNE anomaly.

In summary, the sterile neutrino mass range to explain the MiniBooNE anomaly
is between \SI{100}{eV} and \SI{2.5}{keV}.

\section{The LSND and Reactor Anomalies}
\label{sec:lsnd-reactor}

As shown in \cref{fig:Ue4Um4}, decaying
sterile neutrinos can simultaneously fit the MiniBooNE and LSND anomalies, but
only if cosmological neutrino free-streaming constraints can be avoided
(see discussion under point {\it (6)} above for possible scenarios).
Quantitatively, a parameter goodness-of-fit test \cite{Maltoni:2003cu} reveals
that LSND is incompatible with the rest of the data at the $4.7\sigma$ level
if free-streaming constraints hold.  If the free-streaming problem is solved
by other means, this reduces to $2.1\sigma$, implying consistency.
The best fit to all data including LSND, but excluding free-streaming is found at
$m_4 = \SI{97}{eV}$, $|U_{e4}|^2 = 0.018$, $|U_{\mu 4}|^2 = 0.0015$,
$m_4 \Gamma_4 = \SI{0.87}{eV^2}$, $m_\phi/m_4 = 0.89$.

\begin{figure}
  \centering
  \includegraphics[width=\columnwidth]{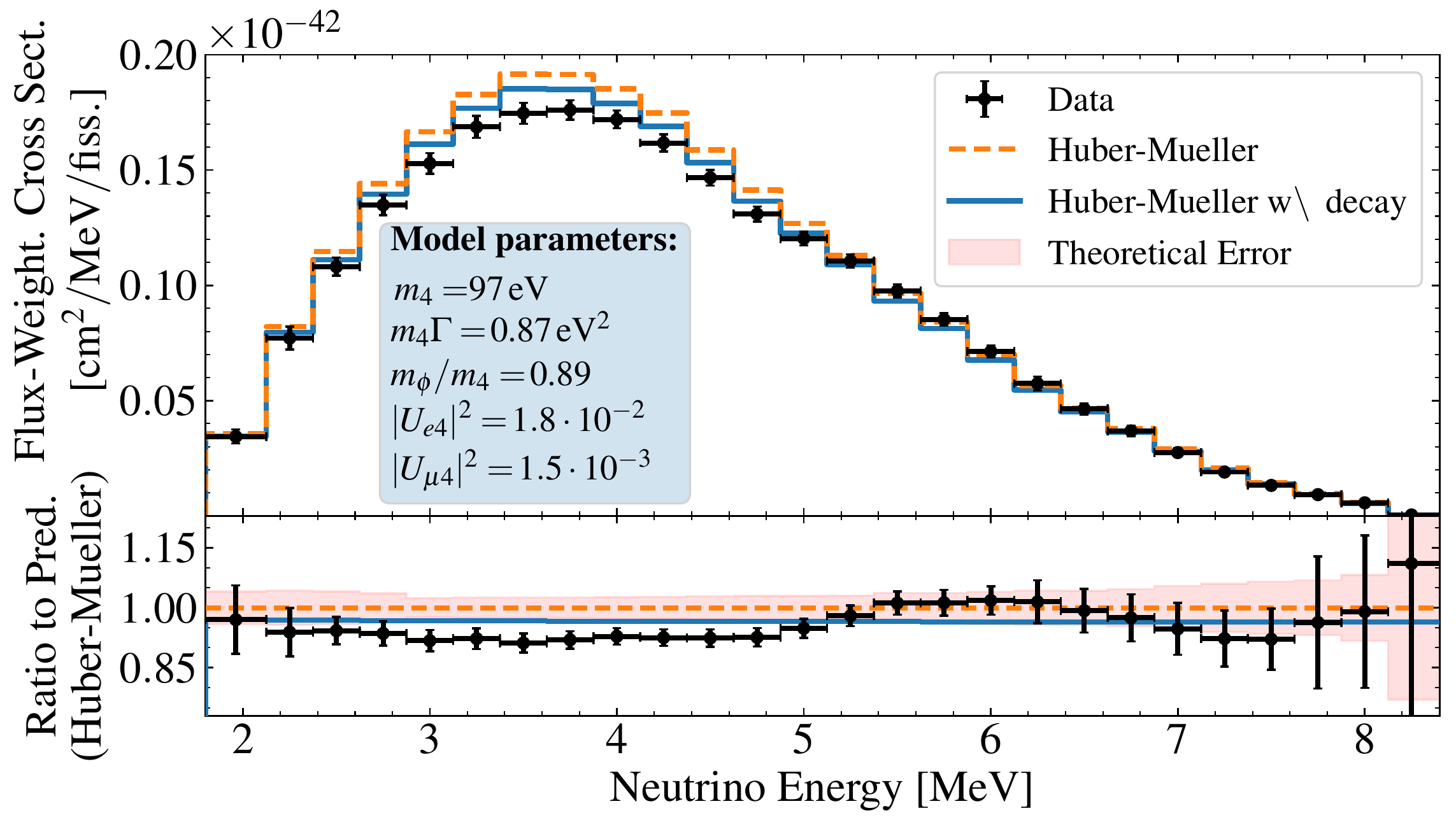}
  \caption{Comparison of the reactor anti-neutrino spectrum predicted
    in the decaying sterile neutrino scenario discussed in this work (blue) to the
    standard Huber--Mueller prediction (orange-dashed) \cite{Mueller:2011nm,Huber:2011wv}
    and to Daya Bay data (black data points with error bars) \cite{An:2016srz}.
    For model parameters motivated by the MiniBooNE and LSND anomalies, a flux
    deficit consistent with the reactor anomaly can be accommodated.
    (See text for details, and for a discussion of how possible cosmological
    constraints can be avoided.)}
  \label{fig:reactor}
\end{figure}

Interestingly, at this value of $|U_{e4}|^2$, the model can also explain the
flux deficit observed in reactor and gallium experiments \cite{Mention:2011rk,
Acero:2007su, Giunti:2010zu, Dentler:2017tkw, Dentler:2018sju, Kostensalo:2019vmv}.
We test our model against reactor data by comparing to Daya Bay's generic
flux-weighted cross section~\cite{An:2016srz}.
To estimate the viable parameter space we perform a chi-square-test using the covariance matrix given in the same reference. In addition we introduce a 2.4\% systematic flux
normalization error corresponding to the theoretical uncertainty, in accordance with
fig.~28 of ref.~\cite{An:2016srz}.
The $|U_{e4}|^2$ region preferred by reactor experiments is included in
\cref{fig:Ue4Um4}, and a comparison of the reactor neutrino spectrum to
our model prediction is shown in \cref{fig:reactor}.

\section{Detailed Investigation of the Parameter Space}
\label{sec:detailed-param-space}

\begin{figure*}
  \begin{tabular}{ccc}
    \includegraphics[width=0.31\textwidth]{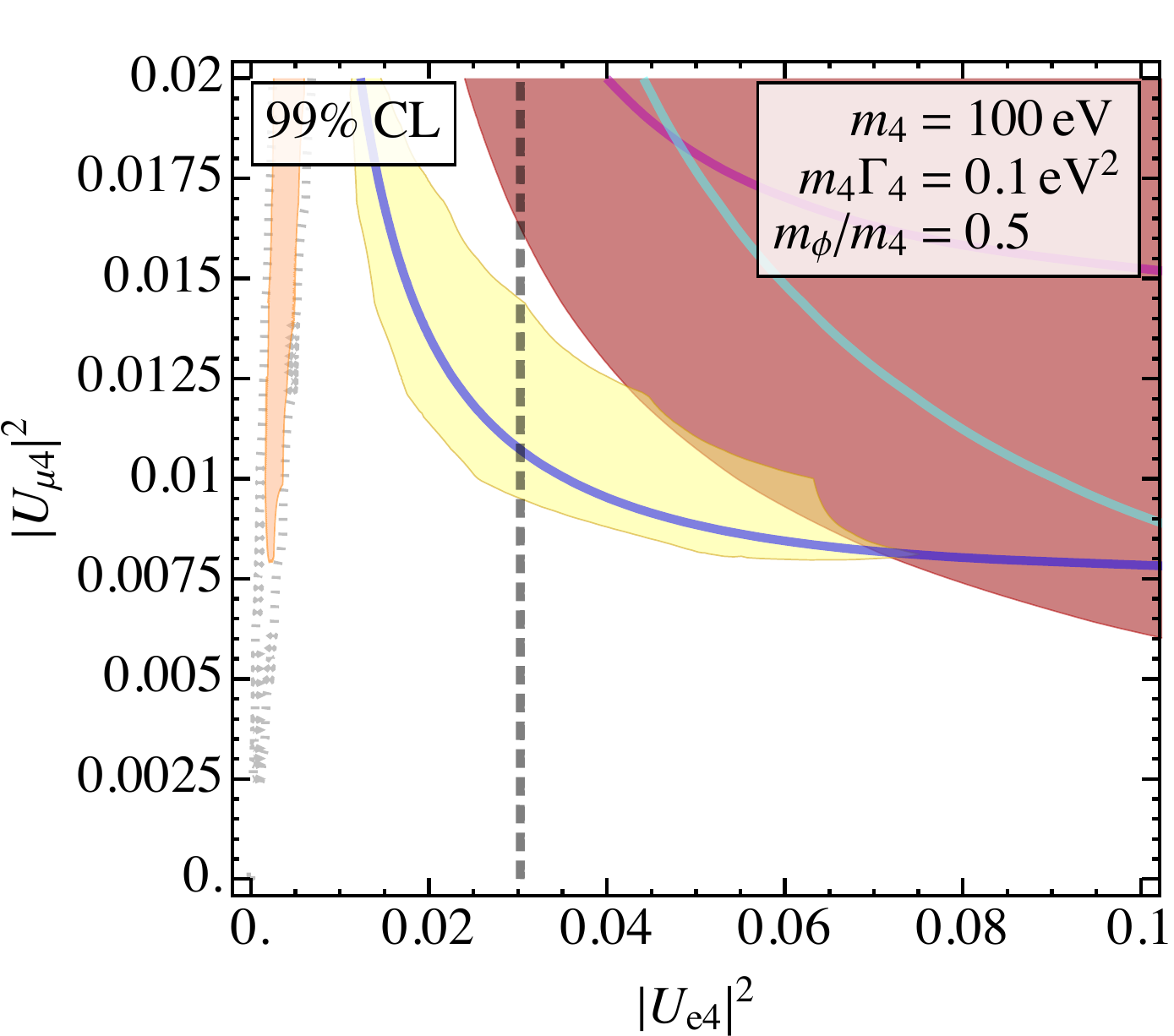}  &
    \includegraphics[width=0.31\textwidth]{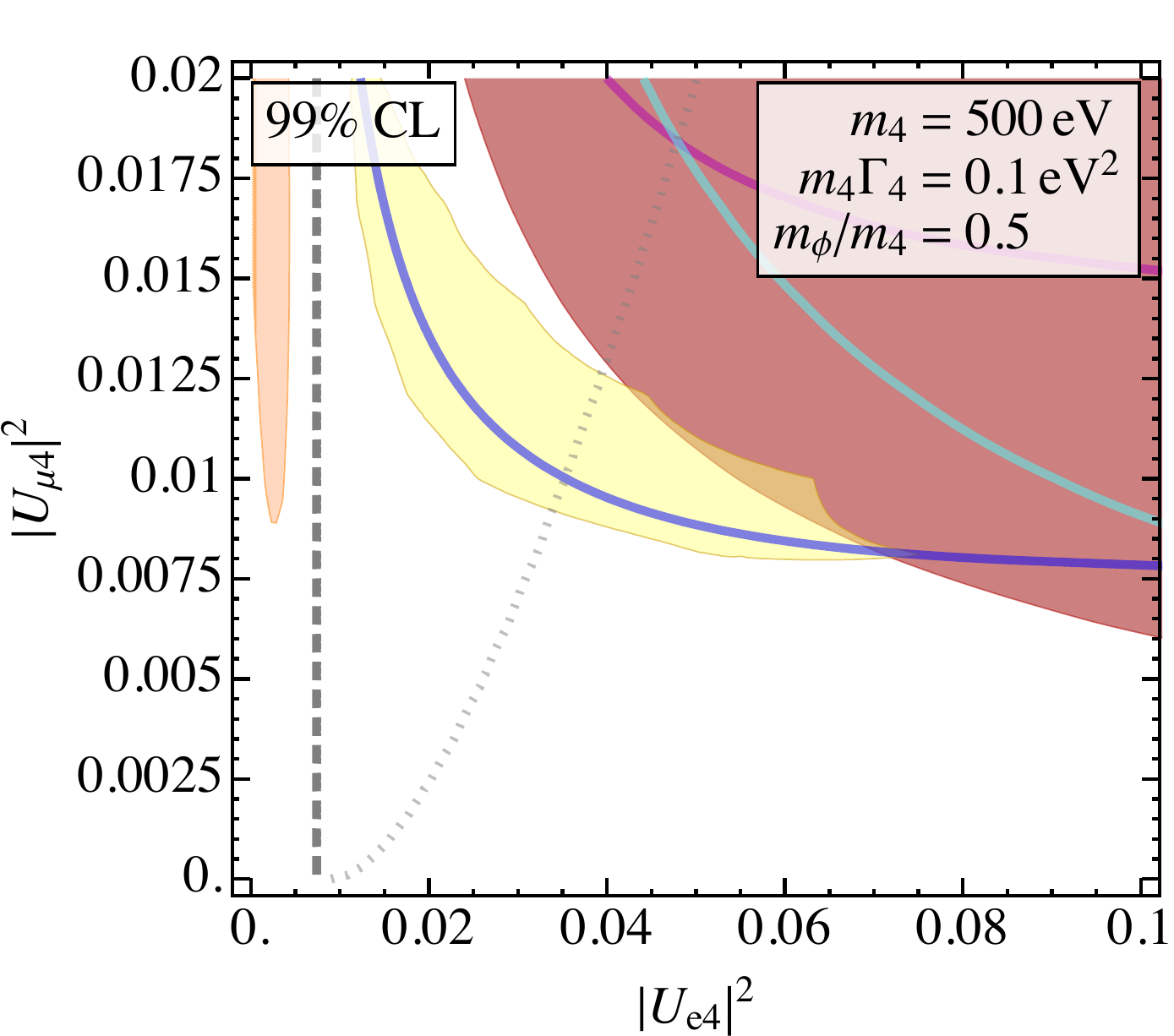}  &
    \includegraphics[width=0.31\textwidth]{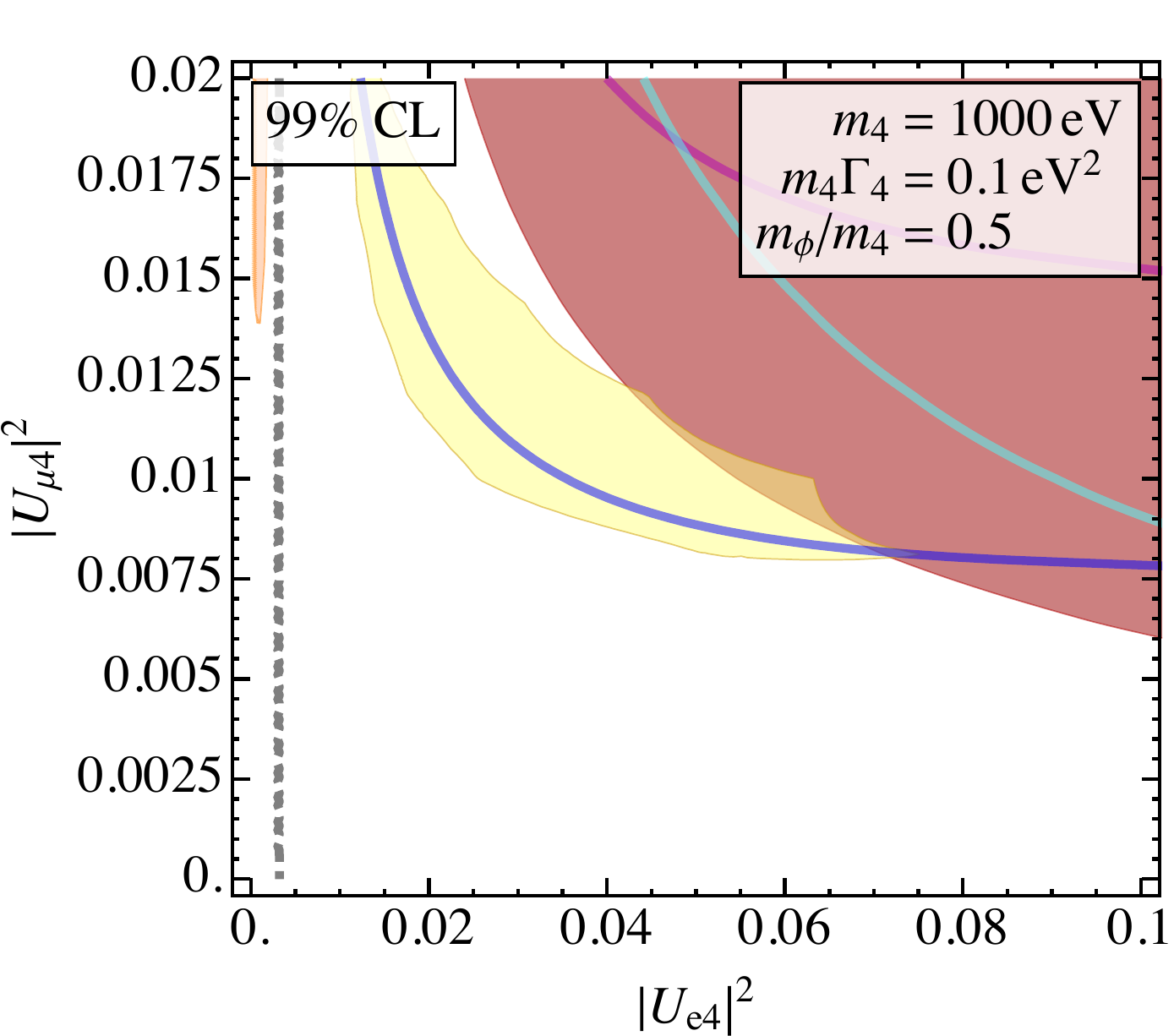} \\
    \includegraphics[width=0.31\textwidth]{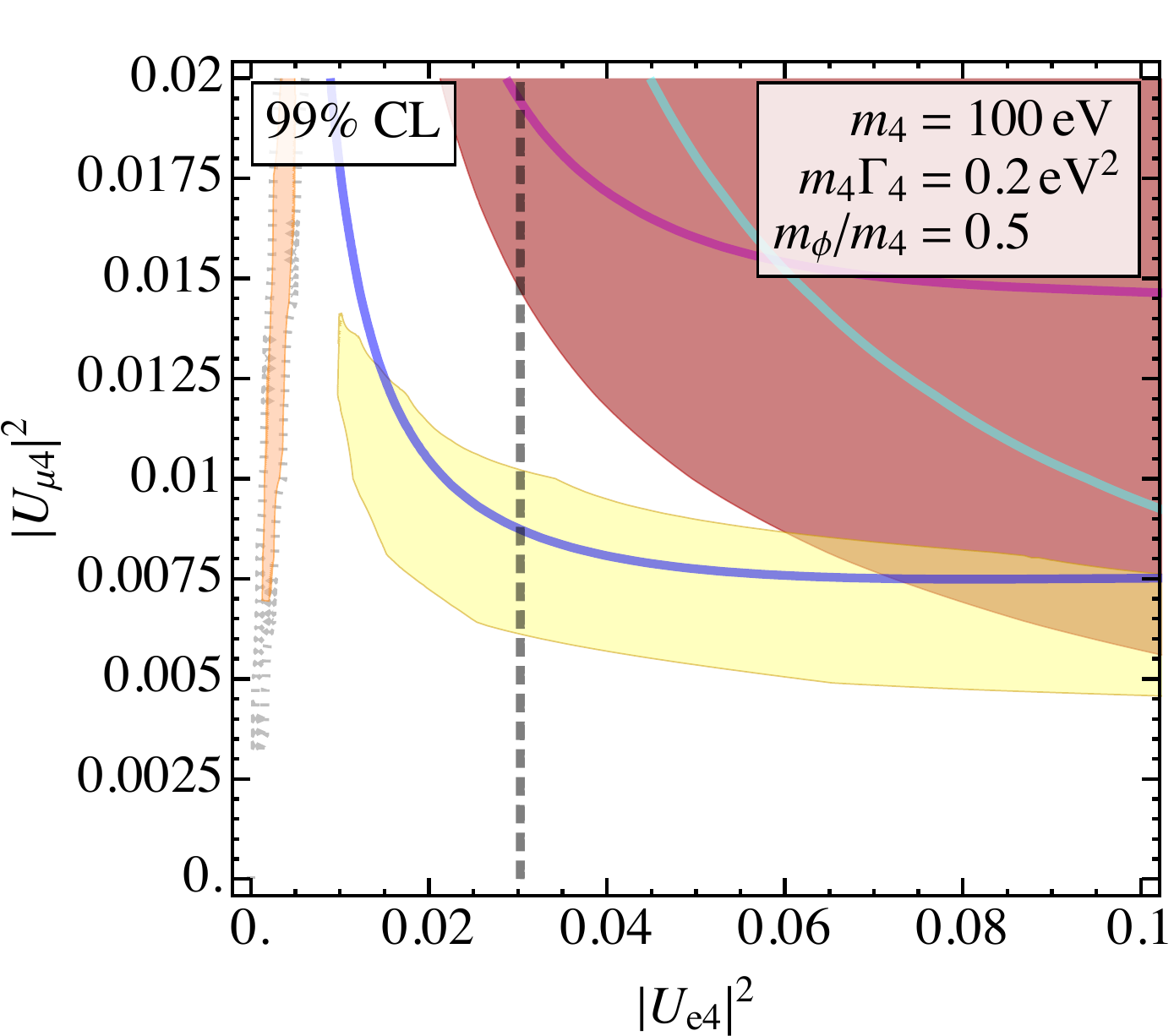}  &
    \includegraphics[width=0.31\textwidth]{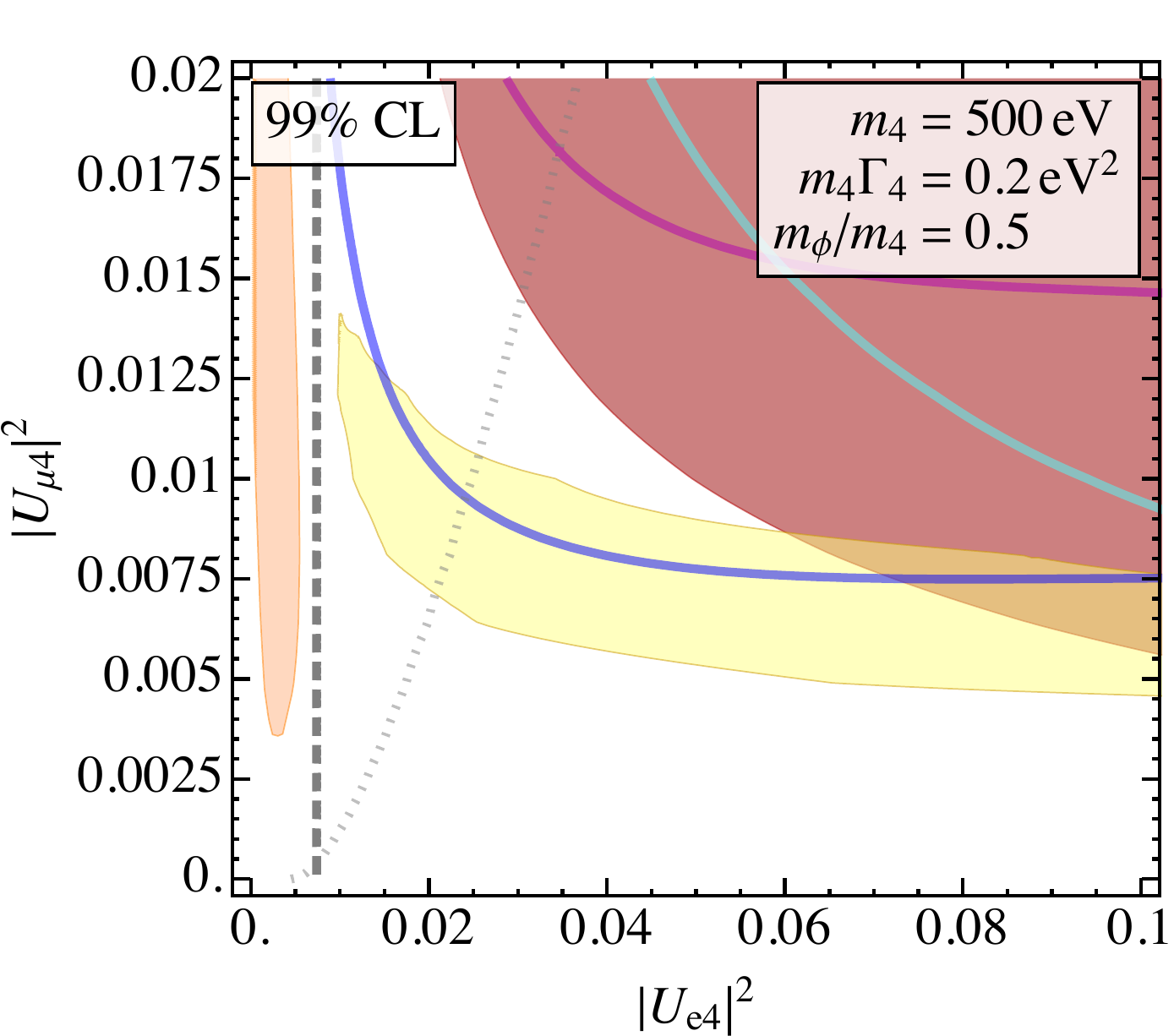}  &
    \includegraphics[width=0.31\textwidth]{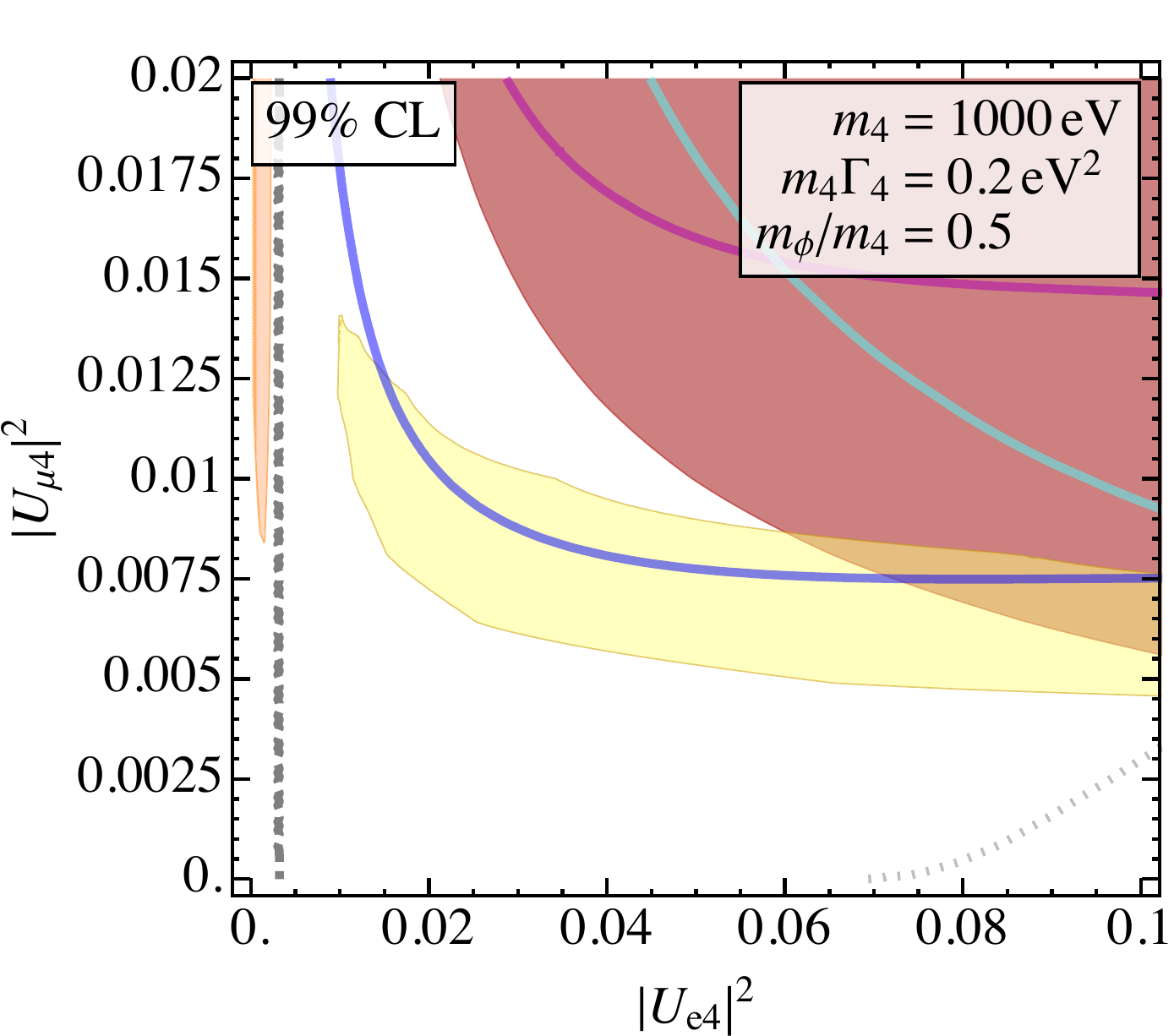} \\
    \includegraphics[width=0.31\textwidth]{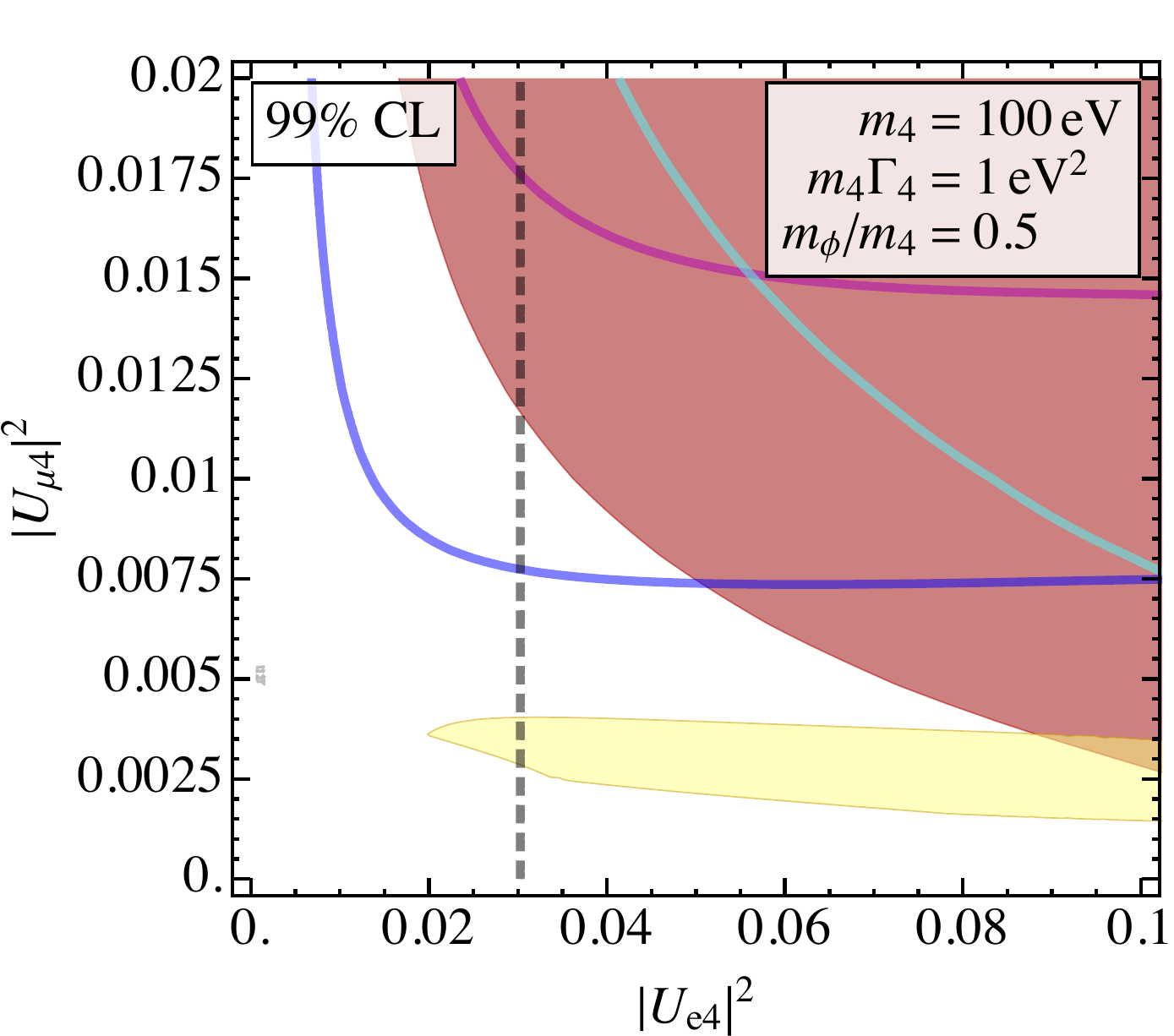}    &
    \includegraphics[width=0.31\textwidth]{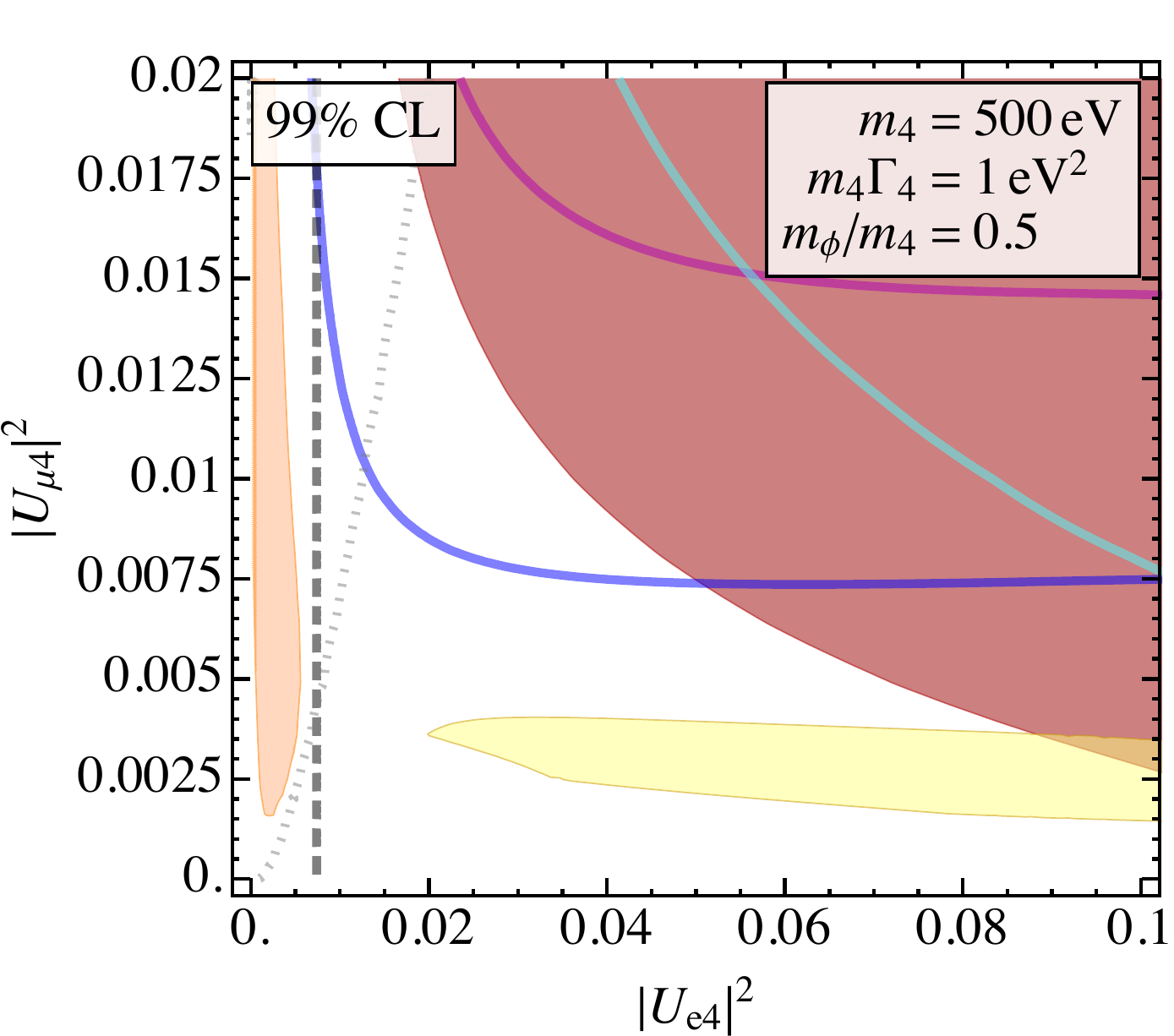}    &
    \includegraphics[width=0.31\textwidth]{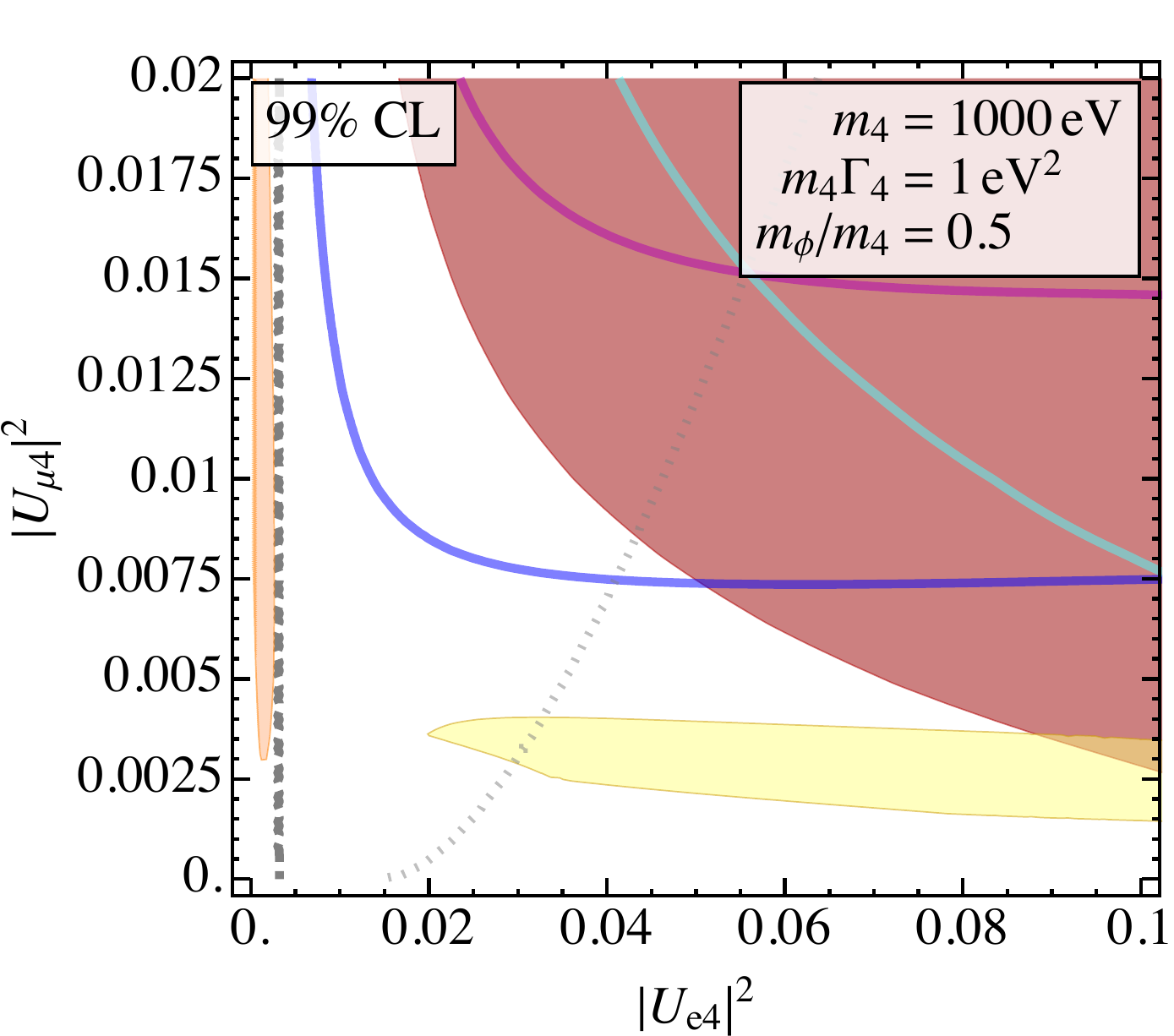}   \\
    \includegraphics[width=0.31\textwidth]{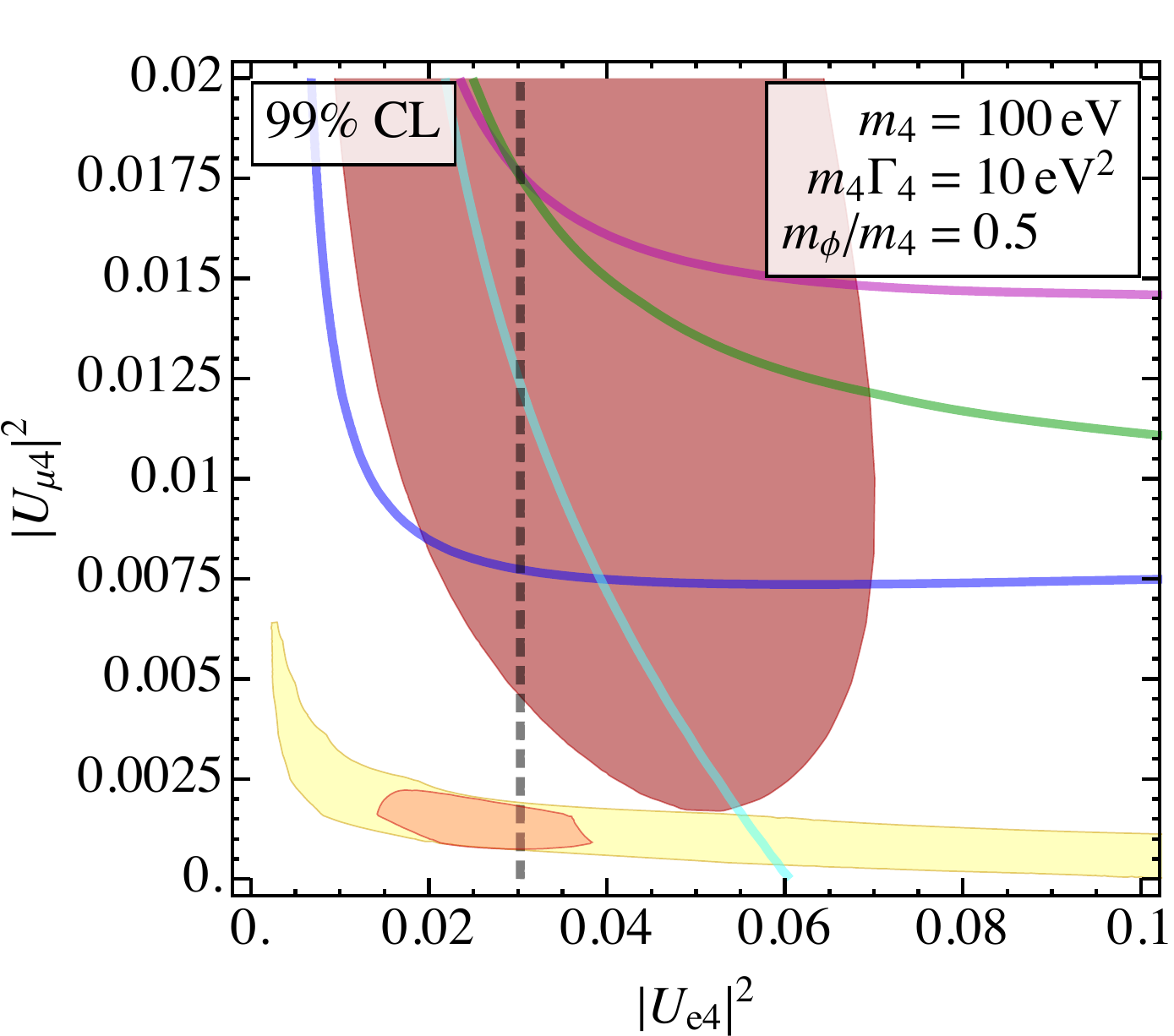}   &
    \includegraphics[width=0.31\textwidth]{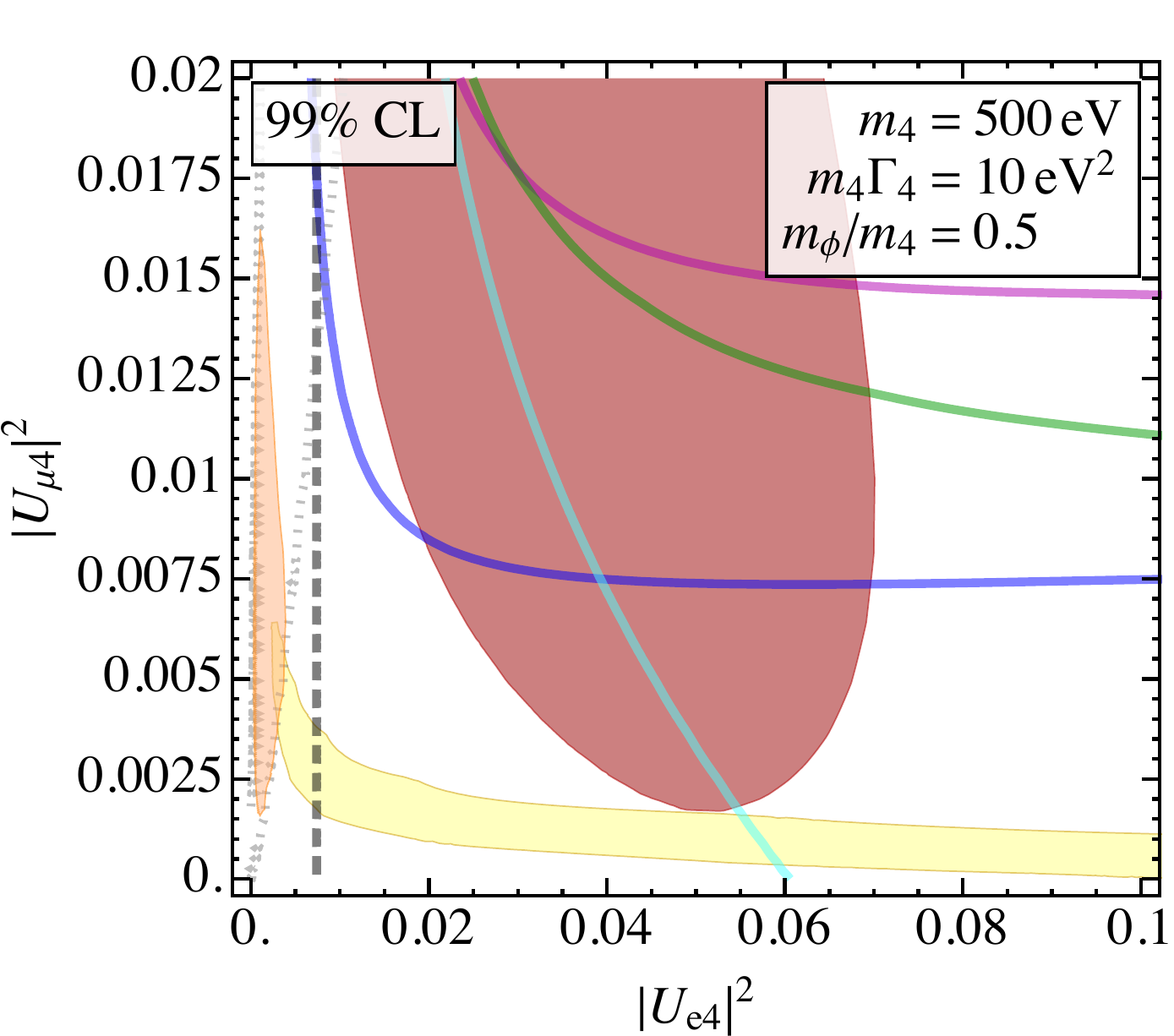}   &
    \includegraphics[width=0.31\textwidth]{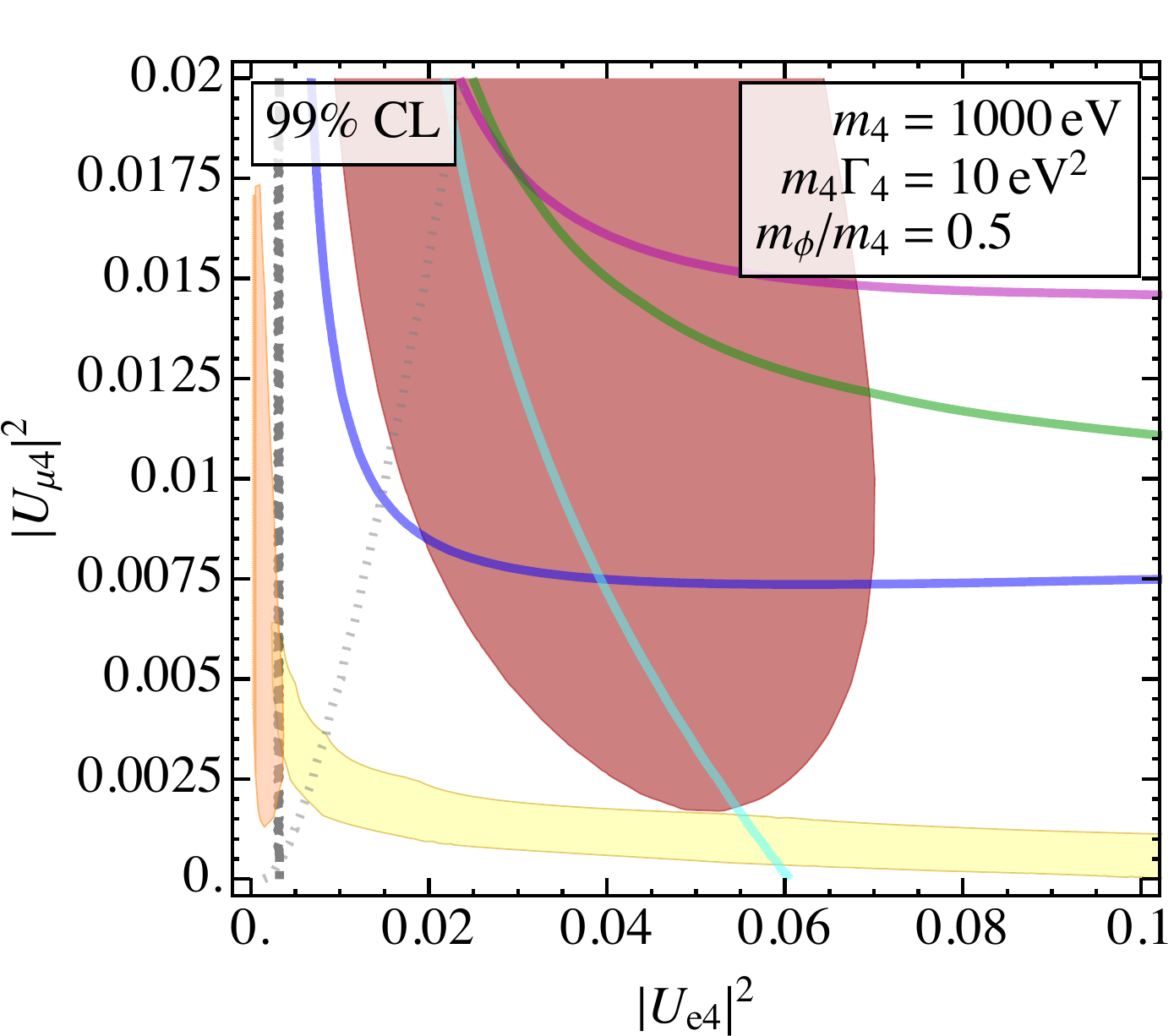}
  \end{tabular}
  \caption{Slices through the 5-dimensional parameter space of decaying sterile
    neutrinos at $m_\phi / m_4 = 0.5$ fixed.
    The color code is the same as in \cref{fig:Ue4Um4}.}
  \label{fig:Ue4Um4-0.5}
\end{figure*}

\begin{figure*}
  \begin{tabular}{ccc}
    \includegraphics[width=0.31\textwidth]{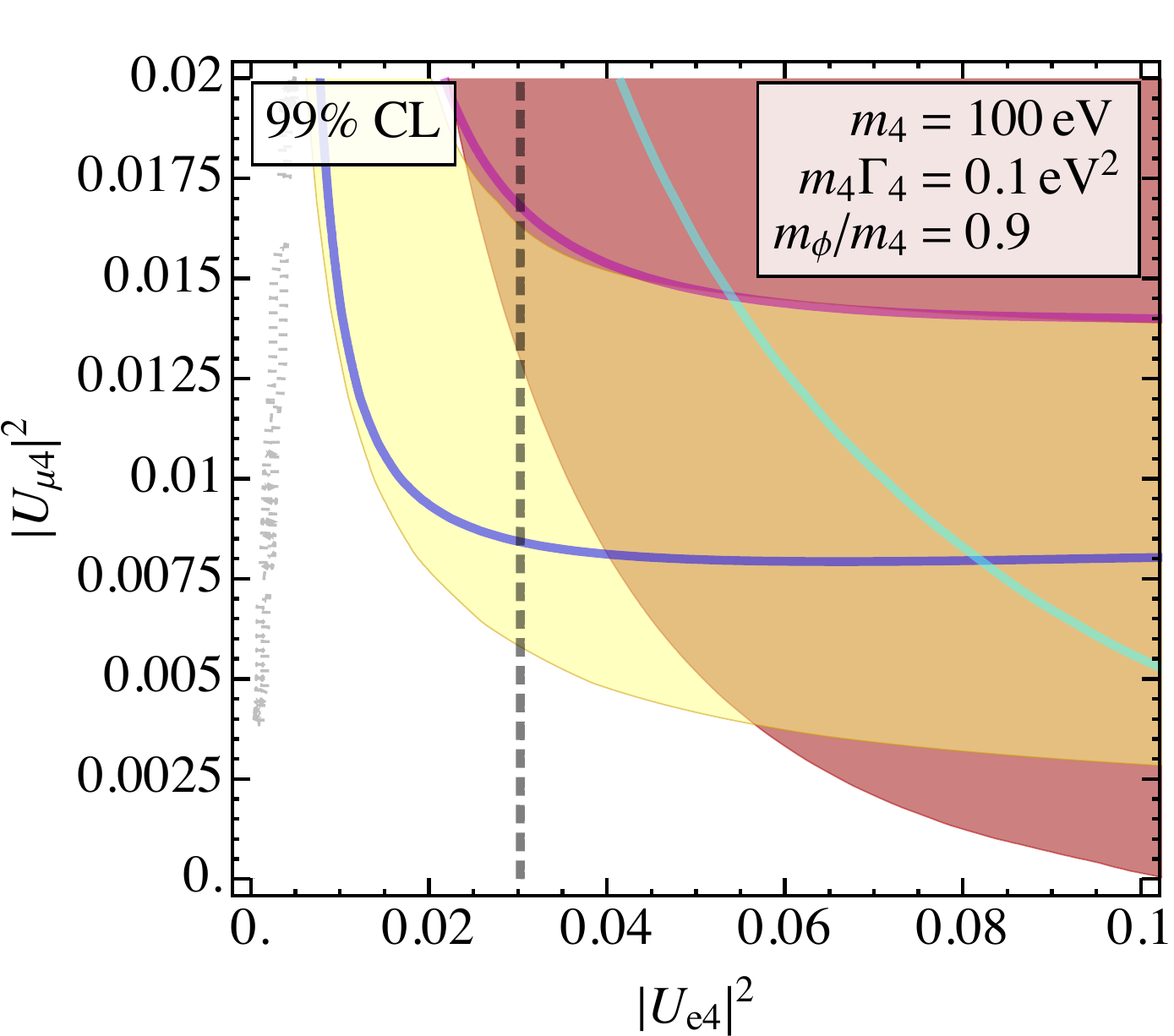}  &
    \includegraphics[width=0.31\textwidth]{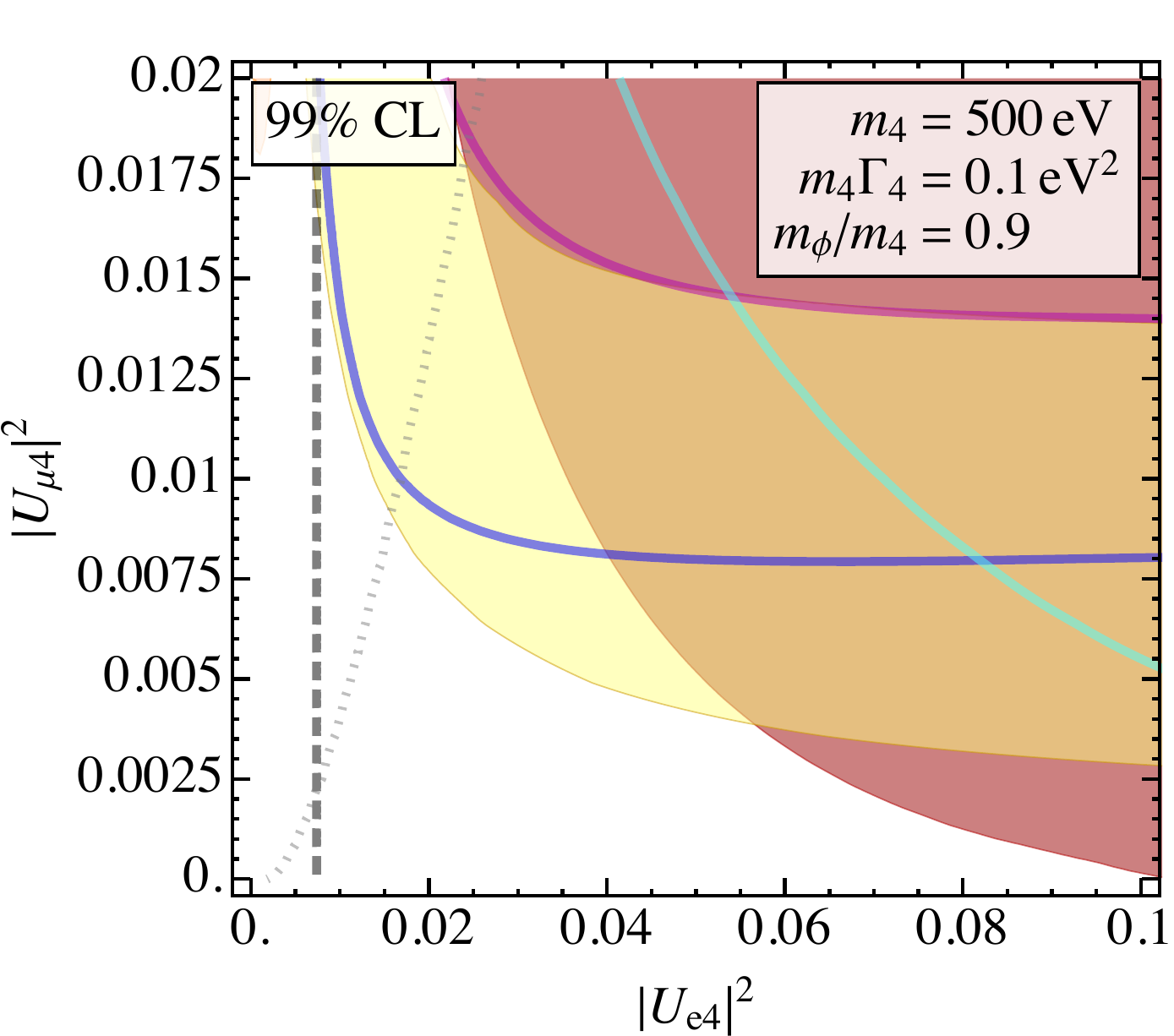}  &
    \includegraphics[width=0.31\textwidth]{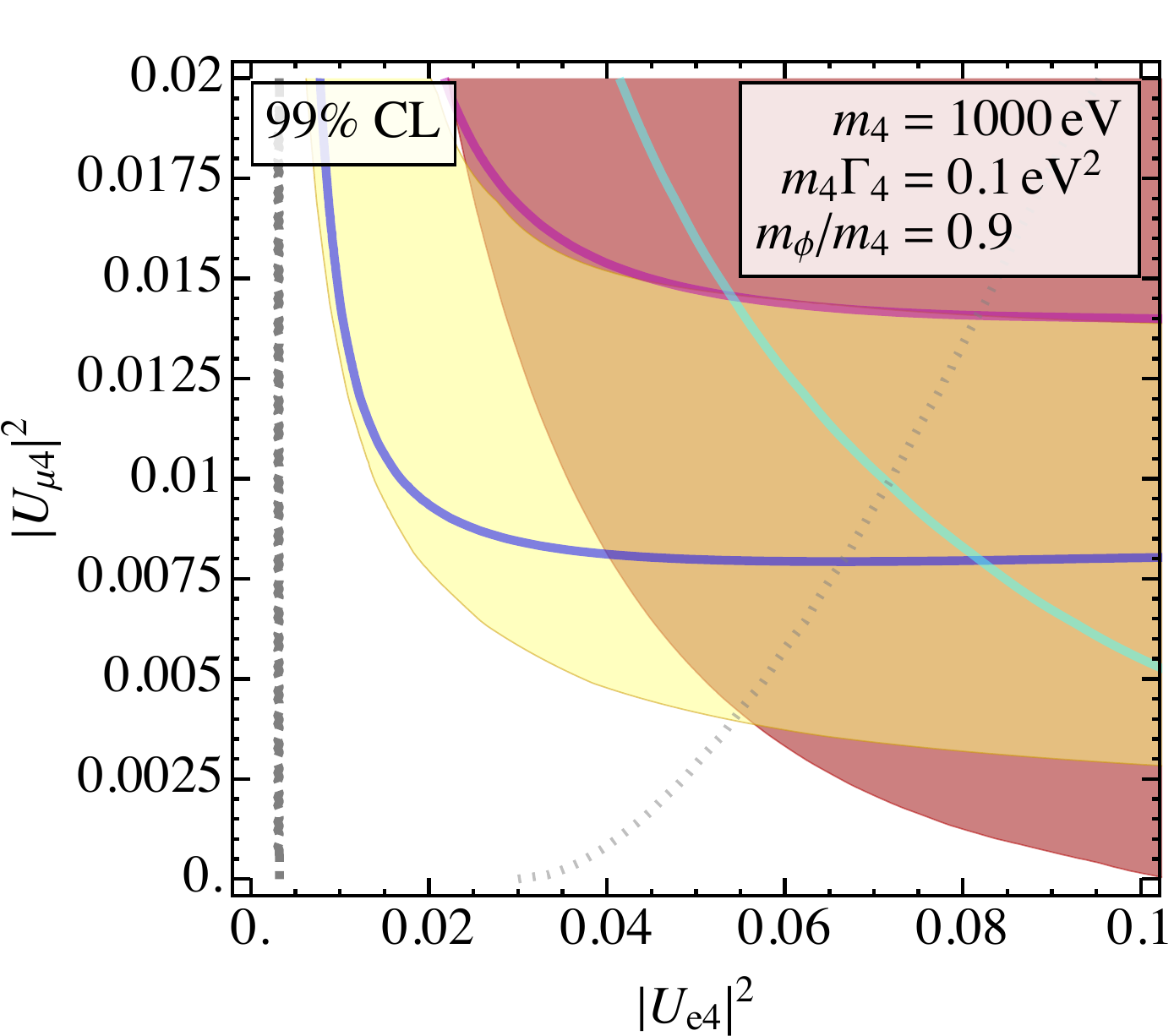} \\
    \includegraphics[width=0.31\textwidth]{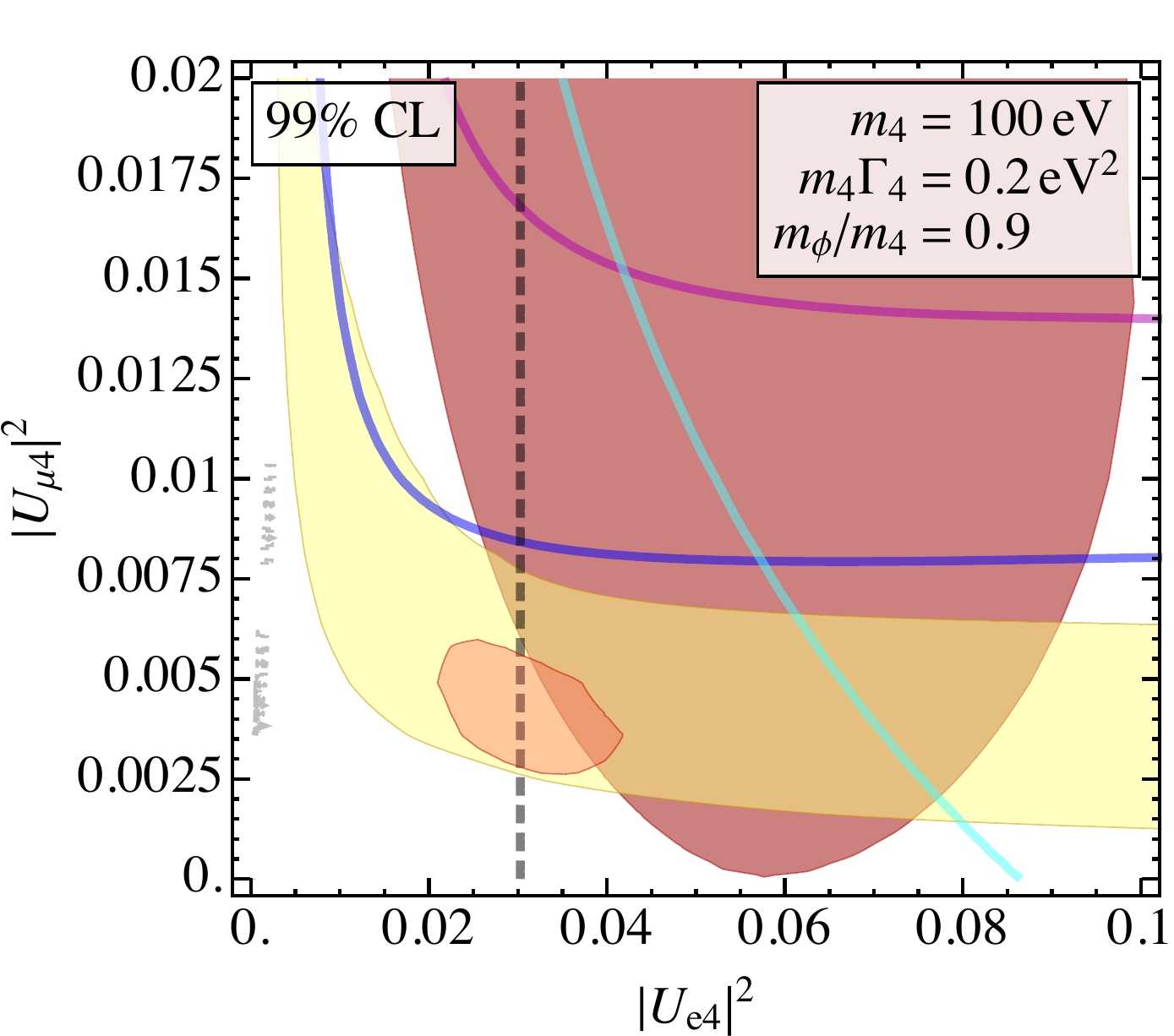}  &
    \includegraphics[width=0.31\textwidth]{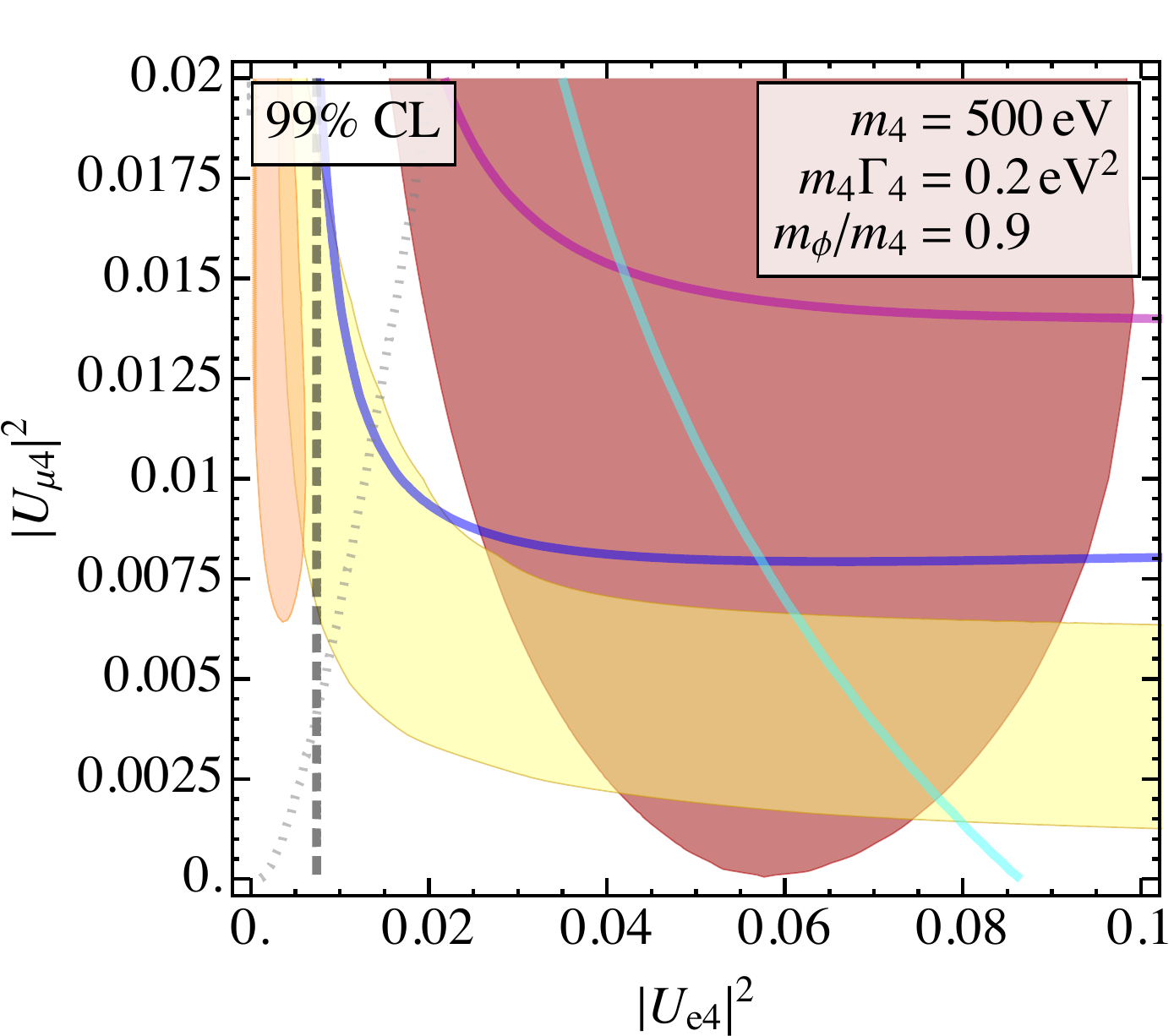}  &
    \includegraphics[width=0.31\textwidth]{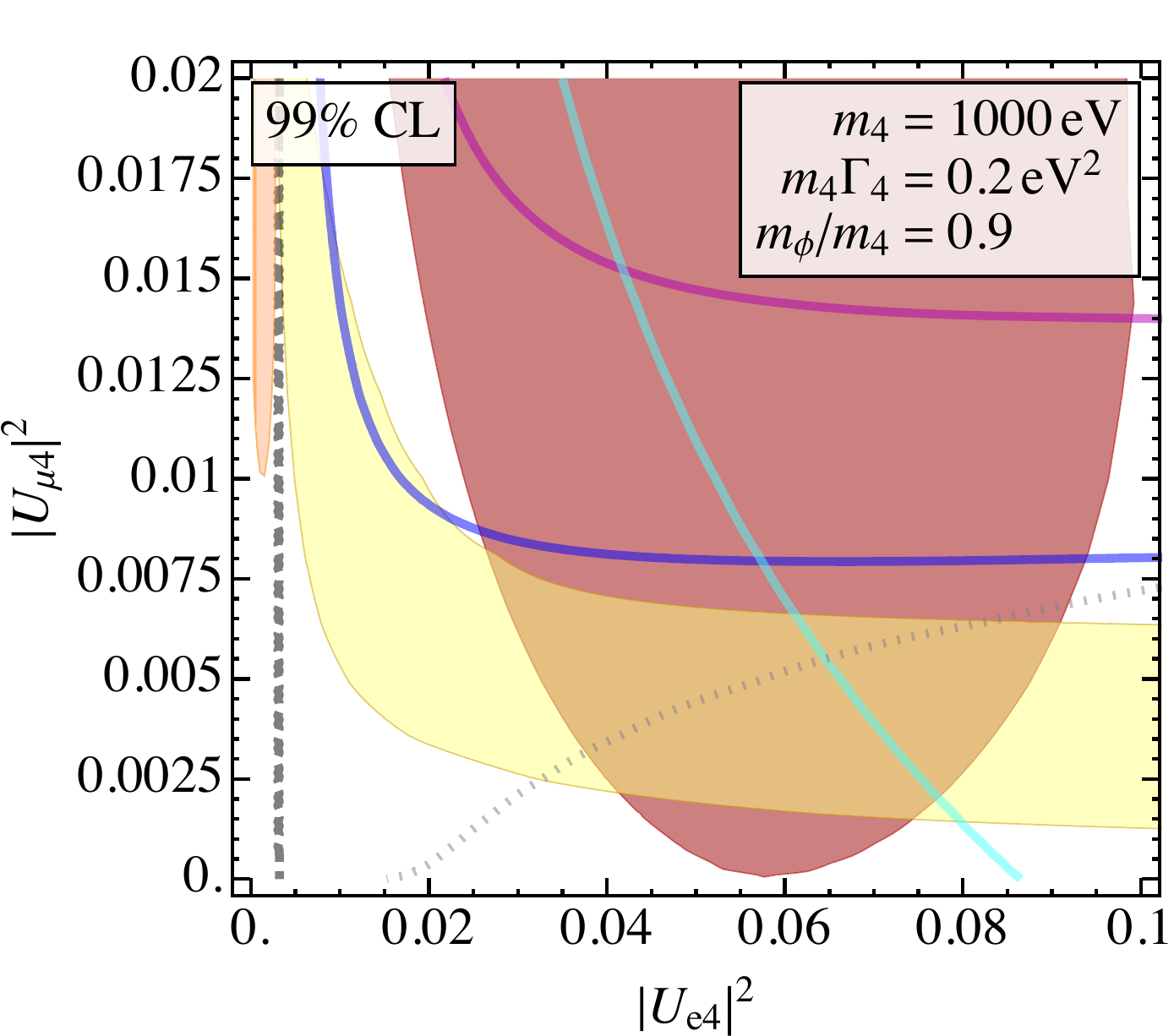} \\
    \includegraphics[width=0.31\textwidth]{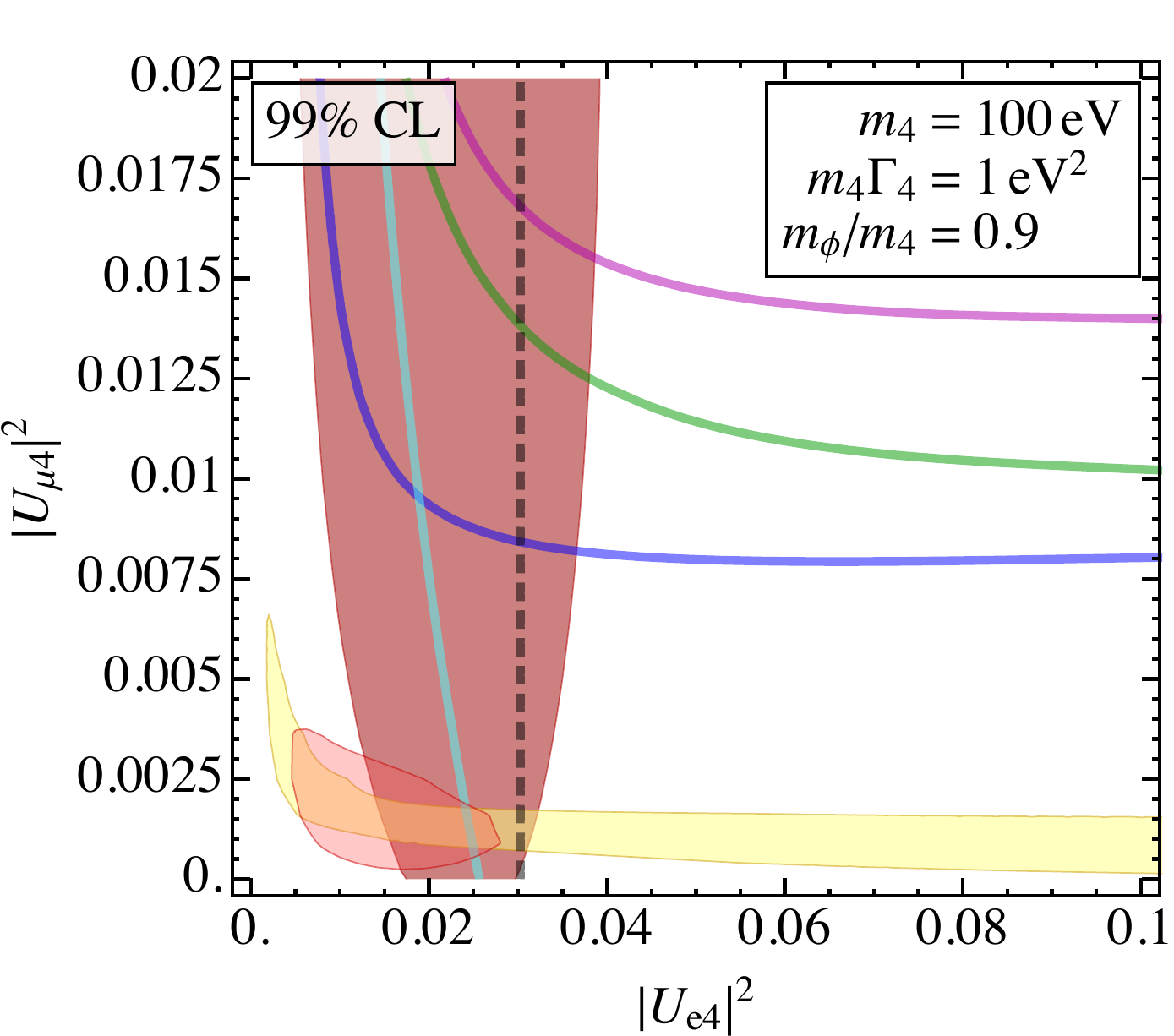}    &
    \includegraphics[width=0.31\textwidth]{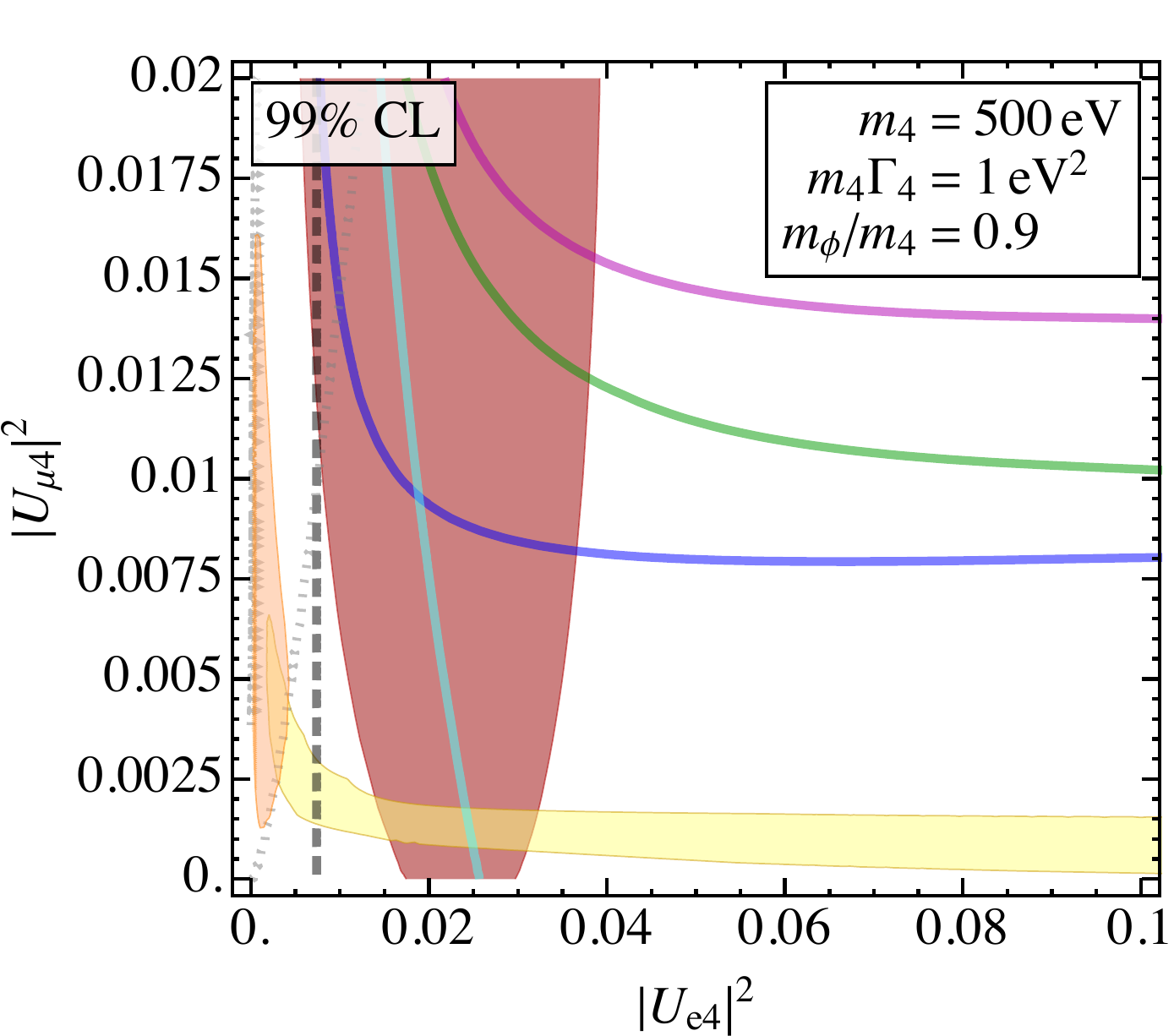}    &
    \includegraphics[width=0.31\textwidth]{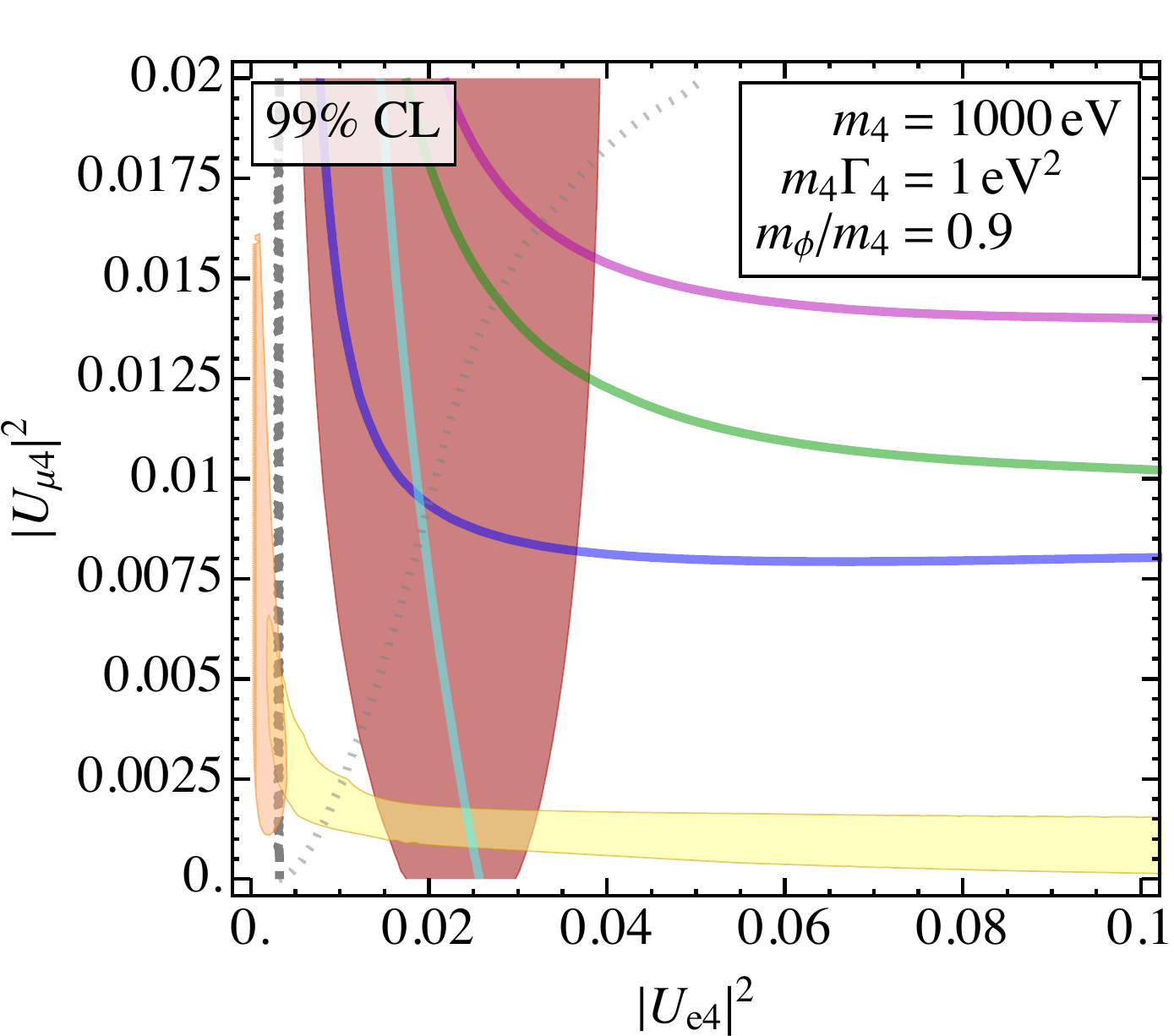}   \\
    \includegraphics[width=0.31\textwidth]{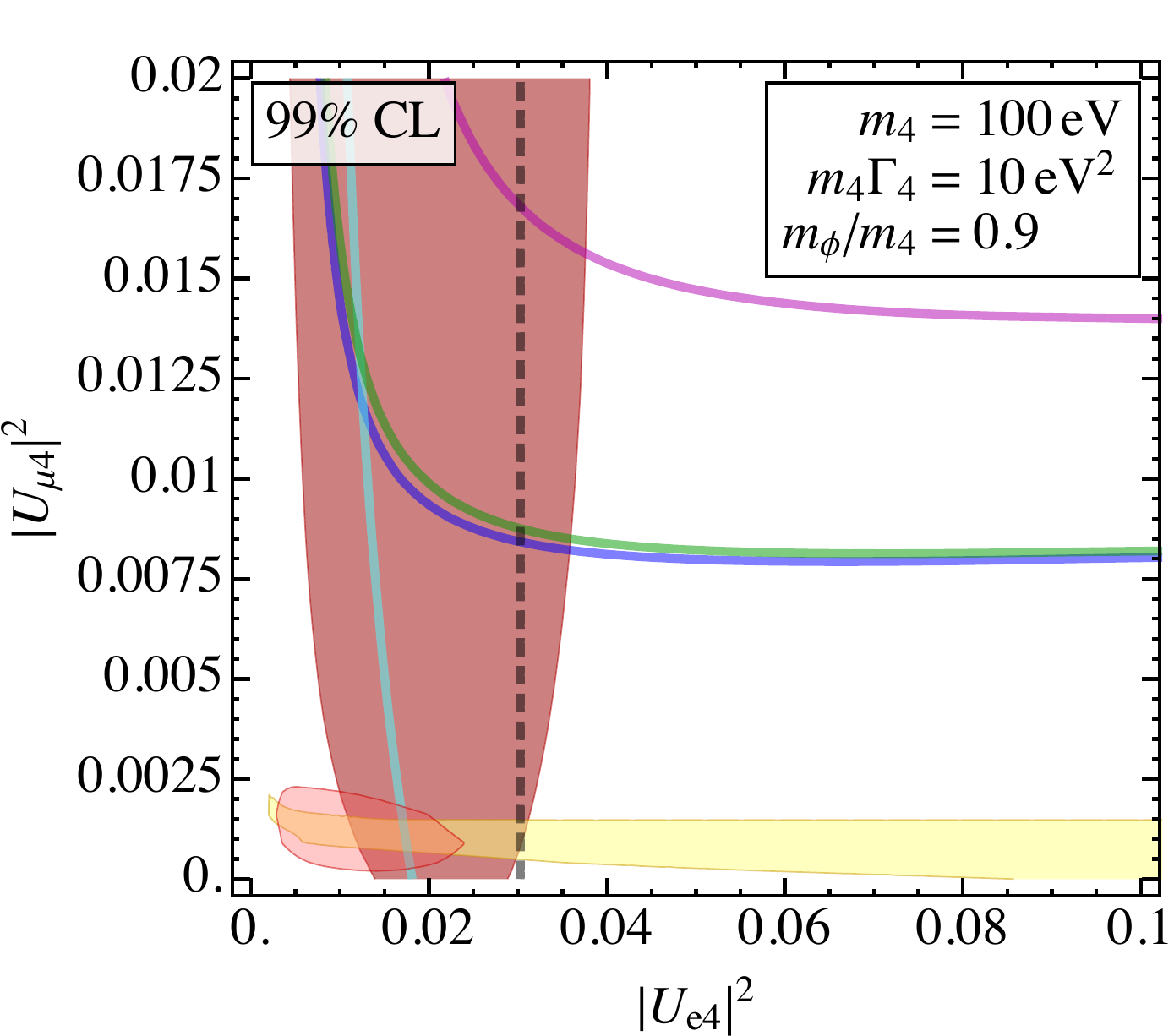}   &
    \includegraphics[width=0.31\textwidth]{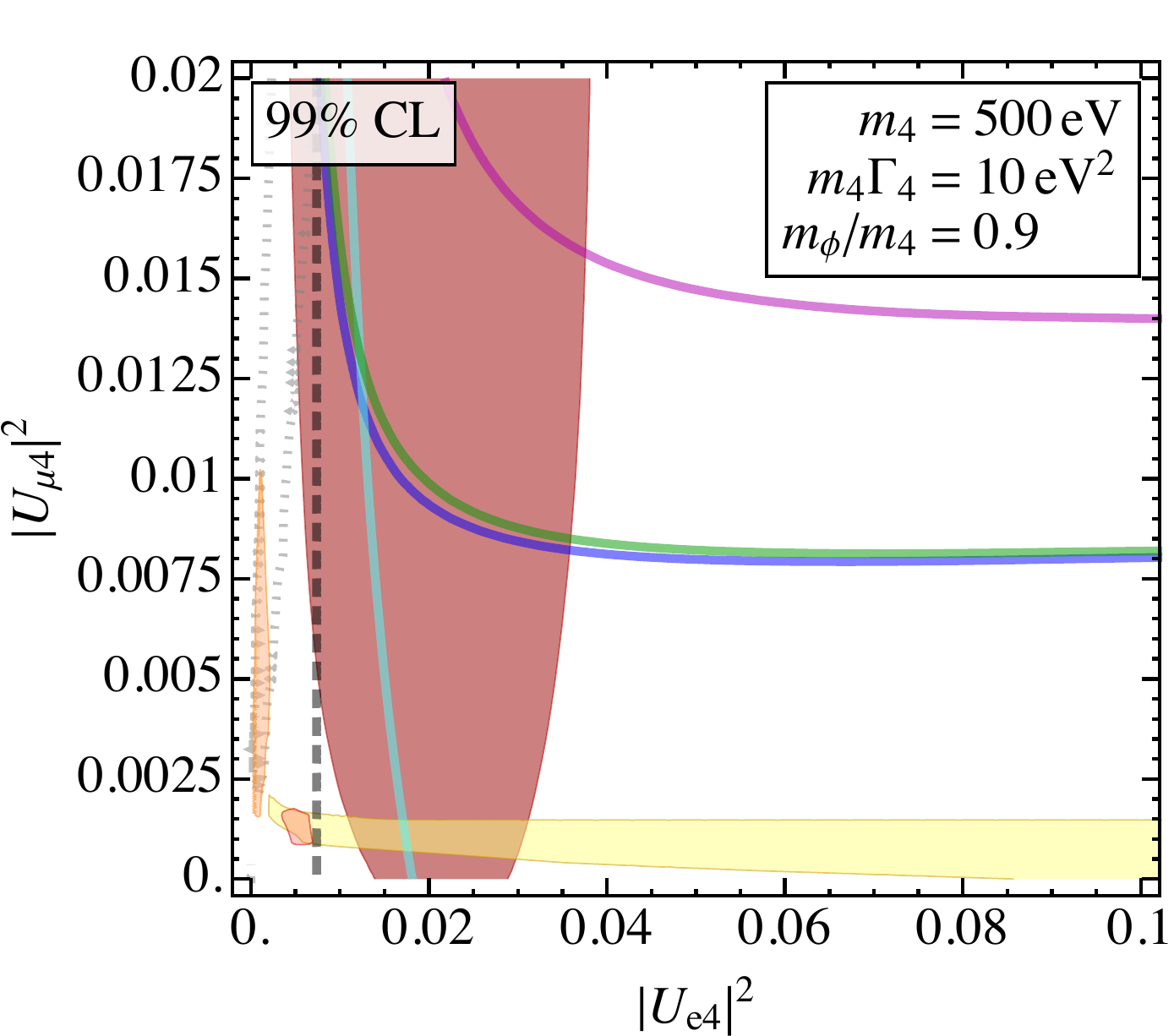}   &
    \includegraphics[width=0.31\textwidth]{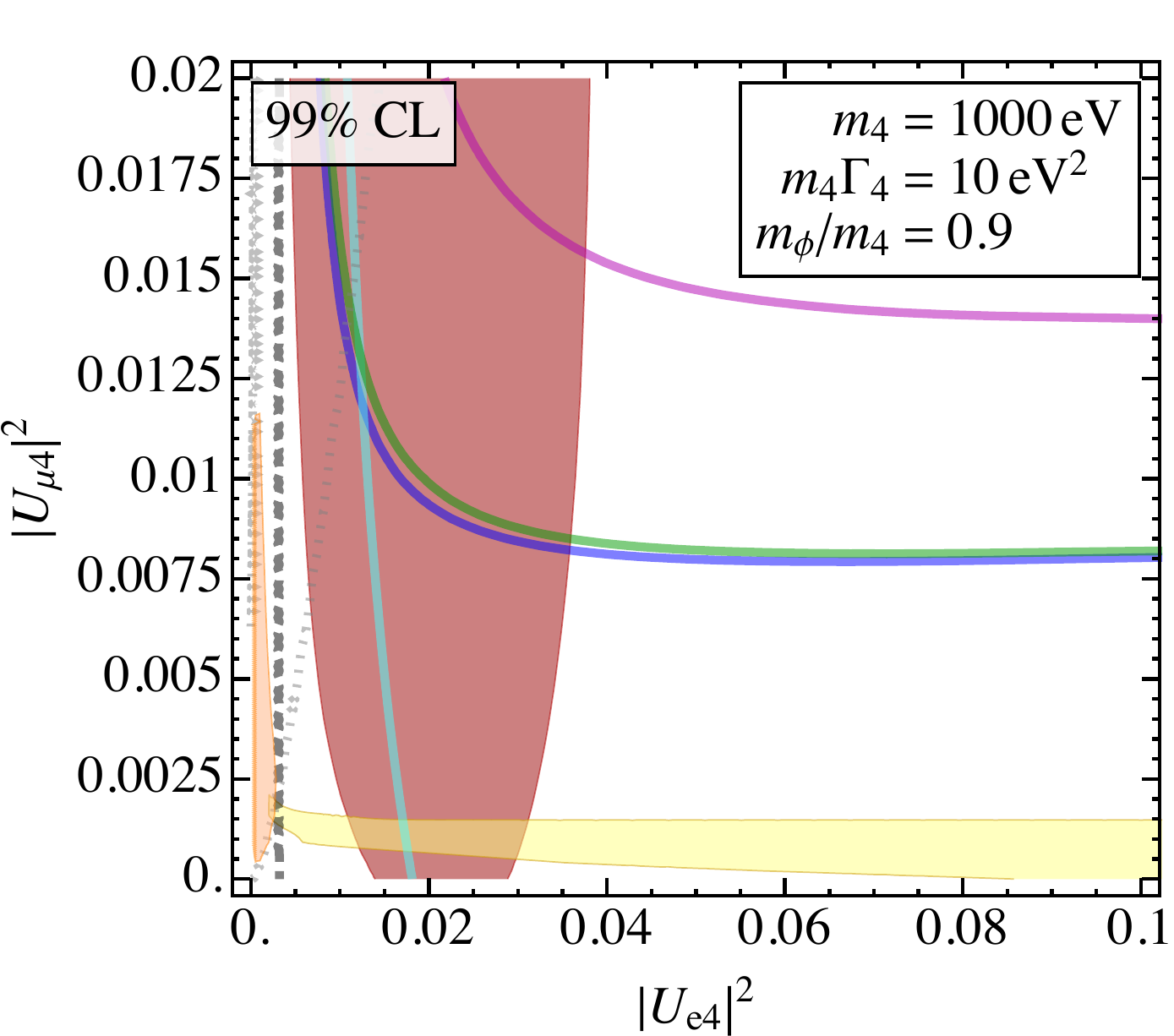}
  \end{tabular}
  \caption{Slices through the 5-dimensional parameter space of decaying sterile
    neutrinos at $m_\phi / m_4 = 0.9$ fixed.
    The color code is the same as in \cref{fig:Ue4Um4}.}
  \label{fig:Ue4Um4-0.9}
\end{figure*}

To supplement \cref{fig:Ue4Um4} and give the reader a broader overview of
the preferred parameter regions of decaying sterile neutrinos, we show in
\cref{fig:Ue4Um4-0.5,fig:Ue4Um4-0.9} additional slices through the
5-dimensional parameter space.

The color coding in the figure is the same as in \cref{fig:Ue4Um4}: the yellow,
banana-shaped regions are preferred by MiniBooNE, the large dark red ones by LSND;
the orange regions at low $|U_{e4}|^2$ correspond to a global fit to MiniBooNE,
OPERA, ICARUS, E776, KARMEN, nuclear beta decay spectra, and cosmological
free-streaming constraints; bright red regions show instead a global fit
to MiniBooNE, LSND, OPERA, ICARUS, E776, KARMEN, and nuclear beta spectra,
but excluding the free-streaming constraint. Solid lines indicate constraints from
OPERA (blue), ICARUS (purple), KARMEN (cyan), E776 (green), nuclear beta decay
spectra (black dashed), and free-streaming in the early Universe (black dotted).
The region to the right of the lines is excluded.

We observe that, at smaller values of $m_4 \Gamma_4$, the allowed parameter regions
from short-baseline oscillations (MiniBooNE, LSND, KARMEN) shift towards
larger values of $|U_{e4}|^2$ and $|U_{\mu 4}|^2$. In this case, only a small
fraction of neutrinos decays before reaching the detector, making the
phenomenology more similar to that of $3+1$ models without decay.
Strong constraints from beta decay spectra and from cosmology imply that
a good global fit cannot be achieved at $m_4 \Gamma_4 \ll \SI{1}{eV^2}$.

Regarding the dependence of the fit on $m_4$, we note that smaller values
of $m_4$ are favored by beta decay spectra, but disfavored by cosmology,
in agreement with \cref{fig:lab-constraints}.  Exclusion limits from oscillation
experiments do not depend on $m_4$ for $m_4 \gg \si{eV}$.

Comparing \cref{fig:Ue4Um4-0.5} with $m_\phi / m_4 = 0.5$ and
\cref{fig:Ue4Um4-0.5} with $m_\phi / m_4 = 0.9$, we see that it becomes in
general more difficult to fit all experiments at smaller $m_\phi / m_4$.  The
reason is that, at small $m_\phi / m_4$, the active neutrinos produced in
$\nu_4$ and $\phi$ decays have a harder spectrum.  This in particular makes it
more difficult to explain the MiniBooNE low-energy excess.  In fact, for even
smaller values of $m_\phi / m_4$, and in particular for nearly massless $\phi$
(as considered in refs.~\cite{PalomaresRuiz:2005vf, Bai:2015ztj}), the
MiniBooNE-preferred region would disappear completely from the plots.

\section{Conclusions}
\label{sec:conclusions}

In summary, we have shown that scenarios in which the SM is
extended by a sterile neutrino that has a decay mode to active neutrinos can
well explain the MiniBooNE anomaly without violating any constraints. An
explanation of the LSND and reactor/gallium anomalies is possible if the model
is extended to avoid constraints on neutrino free-streaming in the early
Universe.  The preferred mass of the sterile neutrino is of order few hundred
eV.

\section*{Acknowledgments}

We would like to thank Andr\'{e} de Gouvea, Frank Deppisch, Matheus Hostert,
Bill Louis, and Michele Maltoni for very helpful discussions.  We are
particularly grateful to Thomas Schwetz for sharing his fitting codes, from
which some of our own codes took inspiration. We also thank Andr\'{e} de Gouvea,
O. L. G. Peres, Suprabh Prakash and G. V. Stenico for sharing their draft on a
similar neutrino decay solution to the SBL anomalies~\cite{deGouvea:2019qre}.
We thank Miguel Escudero and Samuel J. Witte for sharing their cosmological 
constraint.
 IE would like to thank
CERN and Johannes Gutenberg University Mainz for their kind
hospitality.  This work has been funded by the German Research
Foundation (DFG) under Grant Nos.\ EXC-1098, \mbox{KO~4820/1--1},
FOR~2239, GRK~1581, and by the European Research Council (ERC) under
the European Union's Horizon 2020 research and innovation programme
(grant agreement No.\ 637506, ``$\nu$Directions'').  Fermilab is
operated by Fermi Research Alliance, LLC under contract
No. DE-AC02-07CH11359 with the United States Department of Energy.  IE
acknowledges support from the FPU program fellowship FPU15/03697, the EU Networks FP10ITN ELUSIVES (H2020-MSCA-ITN-2015-674896) and INVISIBLES-PLUS 
(H2020-MSCA-RISE-2015-690575), and the MINECO grant FPA2016-76005-C2-1-P.

\appendix
\section{Impact of Oscillations on the Background Prediction in MiniBooNE}
\label{sec:mb-bg}
 
\begin{figure*}
  \centering
  \begin{tabular}{cc}
    \includegraphics[width=0.48\textwidth]{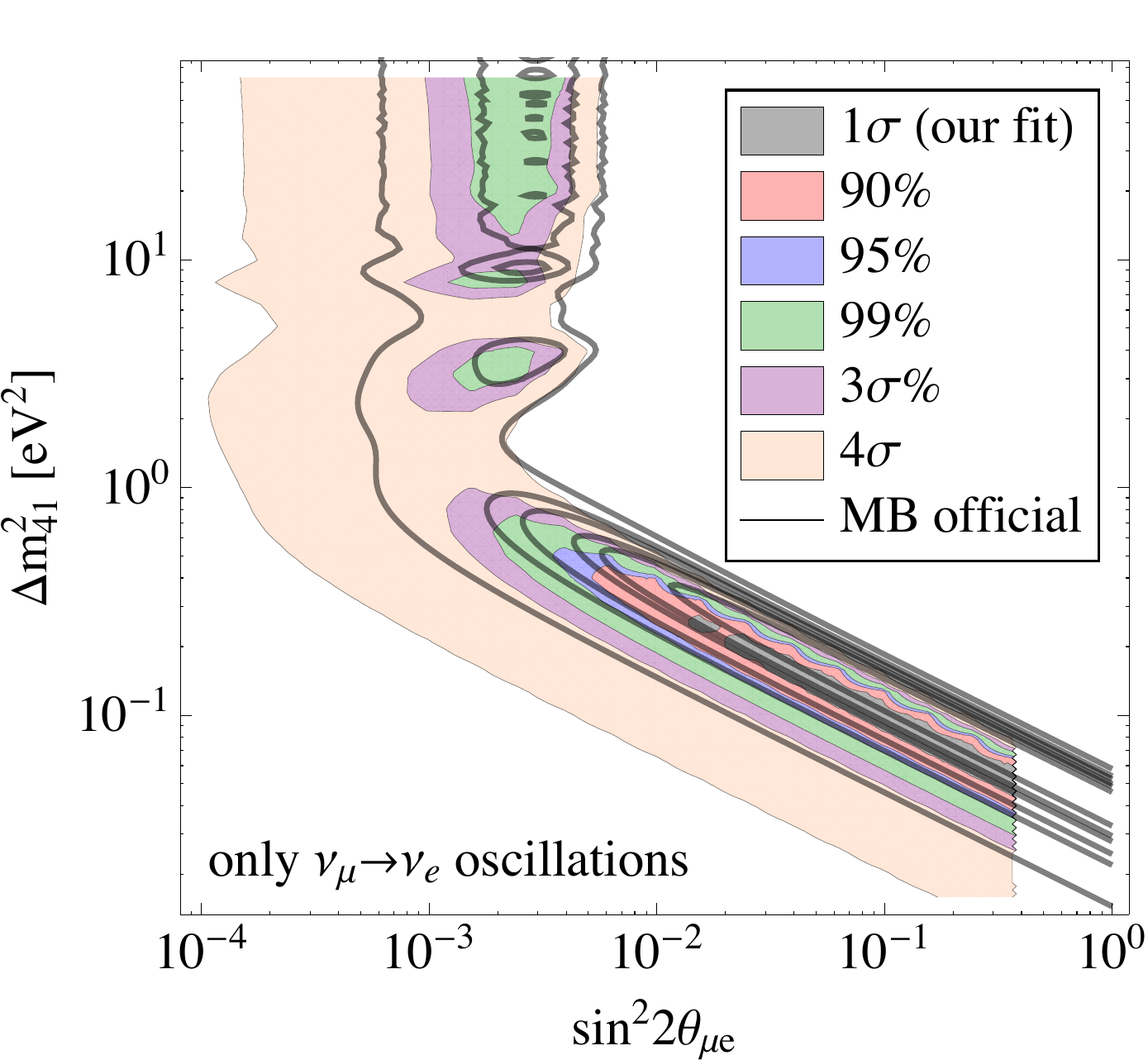} &
    \includegraphics[width=0.48\textwidth]{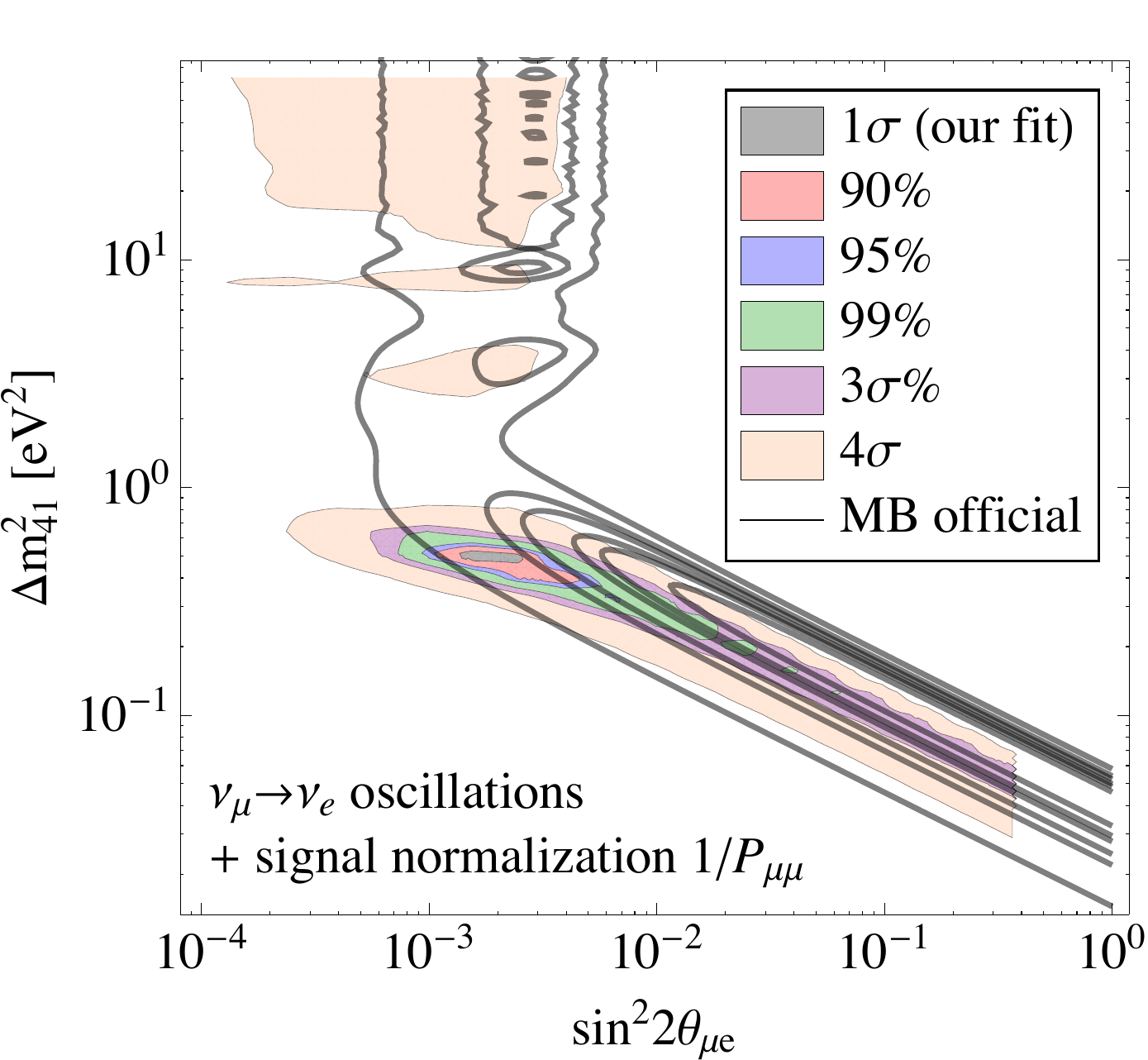} \\
    (a) & (b) \\[0.3cm]
    \includegraphics[width=0.48\textwidth]{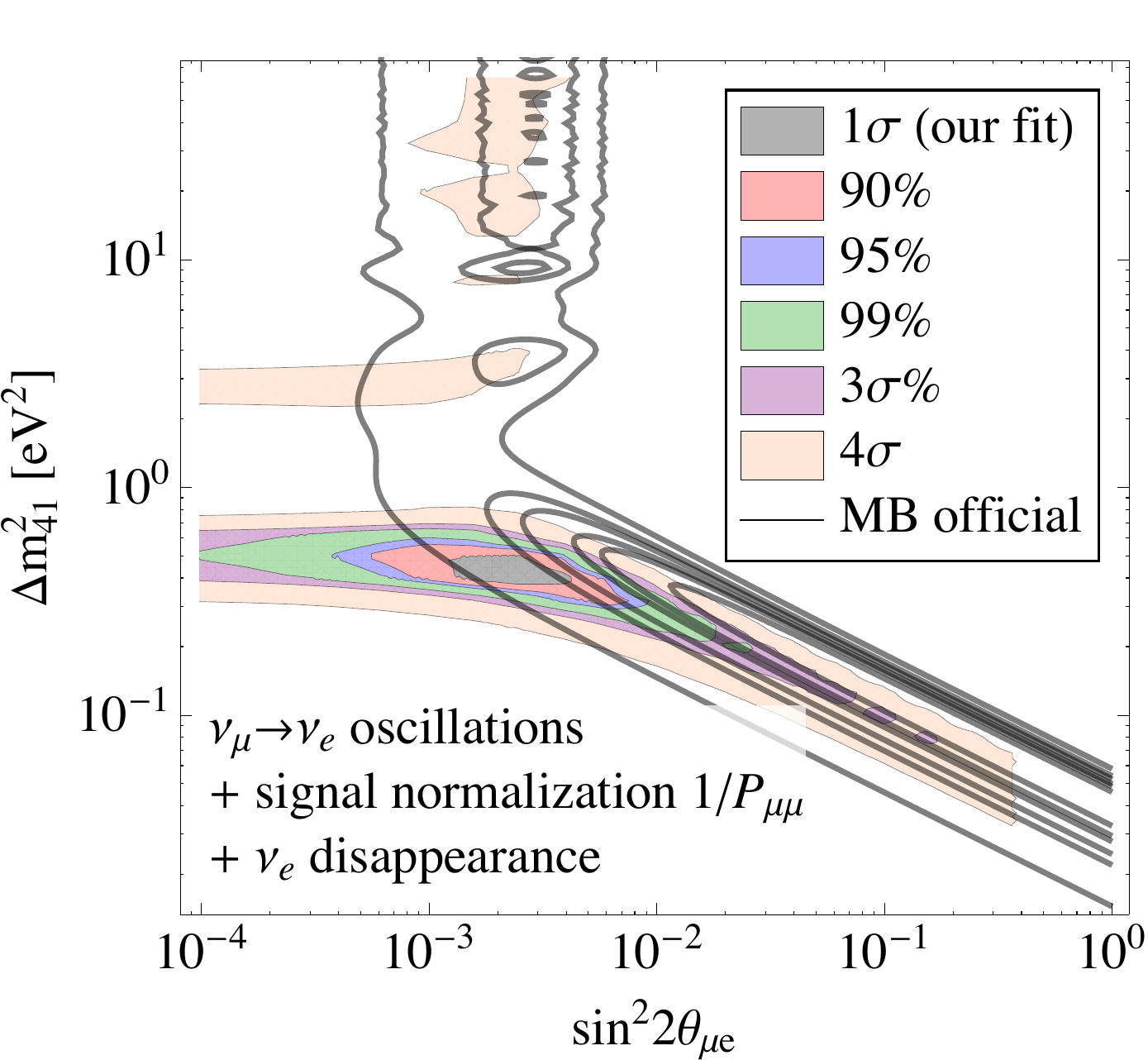} &
    \includegraphics[width=0.48\textwidth]{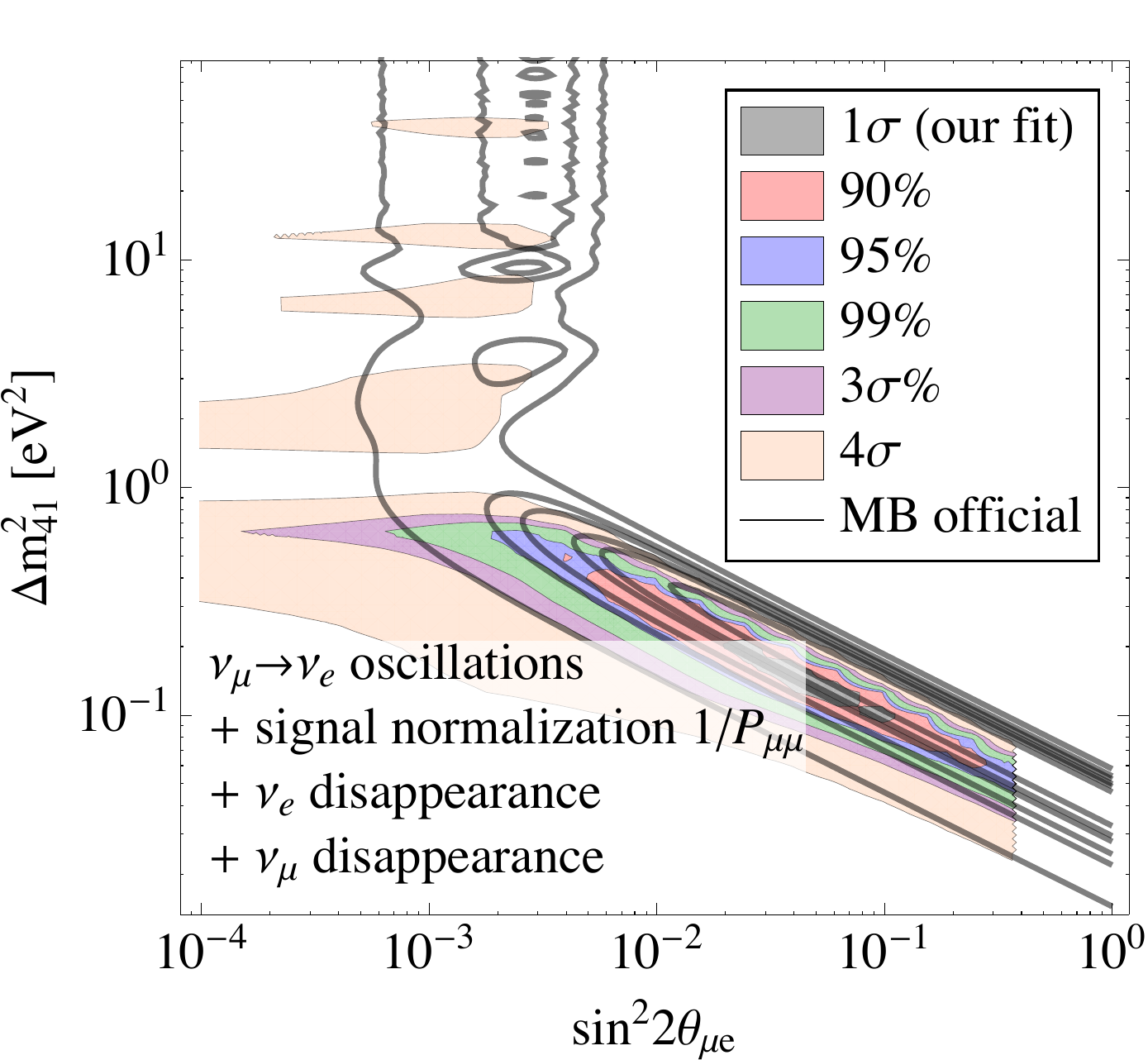} \\
    (c) & (d)
  \end{tabular}
  \caption{Impact of oscillations in the background and control regions on the
    MiniBooNE fit in a simple 3+1 model (oscillations only, no decay). All
    panels show $\Delta m_{41}^2$ vs.\ the effective 2-flavor mixing angle
    $\sin^2 2\theta_{\mu e}$, which in a 3+1 scenario is given by $4 |U_{e4}|^2
    |U_{\mu 4}|^2$.  Panel (a) shows our reproduction (colored regions) of the
    official MiniBooNE fit (black contours), based on the instructions given in
    the supplemental material to ref.~\cite{AguilarArevalo:2010wv} and using
    the data released with ref.~\cite{Aguilar-Arevalo:2018gpe}.  In panel (b),
    we include in addition the impact of $\nu_\mu \to \nu_s$ disappearance on
    the normalization of the signal in each bin.  The colored contours in panel
    (c) include on top of this the effect of $\nu_s$ disappearance on the
    intrinsic $\nu_e$ contamination in the beam.  Panel (d) finally shows the
    additional impact of $\nu_\mu$ disappearance on the sample of $\nu_\mu$
    events that is included in the fit along with the $\nu_e$ sample.  In all
    panels, we show projections of the three-dimensional parameter space spanned
    by $\Delta m_{41}^2$, $|U_{e4}|^2$, and $U_{\mu 4}|^2$ onto the
    $\Delta m_{41}^2$--$\sin^2 2\theta_{\mu e}$ plane, imposing the constraint
    $|U_{e4}|^2 < 0.2$ due to bounds from reactor neutrino experiments.
  }
  \label{fig:bg-osc}
\end{figure*}

In this appendix, we briefly discuss our fit to MiniBooNE data, and in what
ways it differs from the collaborations' fit as described in the supplemental material
to ref.~\cite{AguilarArevalo:2010wv}, and using the data released with
ref.~\cite{Aguilar-Arevalo:2018gpe}. In particular, we consider the following
three effects, which are relevant in a fit to a 3+1 scenario, but are
not encountered in a 2-flavor fit.
\begin{enumerate}
  \item {\bf Normalization of the $\nu_\mu \to \nu_e$ oscillation signal.}
    To predict the number of expected $\nu_e$ events from $\nu_\mu \to \nu_e$
    oscillations for a given set of oscillation parameters, the initial
    $\nu_\mu$ flux must be known.  It is obtained in situ using MiniBooNE's own
    sample of $\nu_\mu$ events.  Note, however, that in a 3+1 model, the
    measured $\nu_\mu$ flux will be reduced by an amount $\sim |U_{\mu 4}|^2$
    due to $\nu_\mu \to \nu_s$ oscillations.  (This effect is unimportant
    in a 2-flavor model, where the deficit is only of order $\sin^2 2\theta_{\mu e}$,
    where $\theta_{\mu e}$ is the effective 2-flavor mixing angle.)
    We account for this effect by first computing the expected $\nu_e$ signal
    based on the unoscillated MiniBooNE flux, and then diving it by the
    $\nu_\mu$ survival probability in each bin.

    The impact of this change in normalization is illustrated in the top
    panels of \cref{fig:bg-osc}.  The colored region in panel (a) of this figure shows our
    reproduction of the official MiniBooNE fit, which is shown as black contours.
    In panel (b), we have included the change in normalization for the signal.

  \item {\bf Oscillations of the $\nu_e$ backgrounds.} Part of the MiniBooNE
    background is constituted by the intrinsic $\nu_e$ contamination in the beam.
    In a 2-flavor fit, this contribution to the total event rate is only
    modified by a factor of order $\sin^2 2\theta_{\mu e}$, but in the full
    3+1 framework, it is reduced by a factor of order $|U_{e4}|^2$ instead.
    The impact of this modification to the background sample is shown in 
    \cref{fig:bg-osc} (c).

  \item {\bf Oscillations of the $\nu_\mu$ sample.} The fit
    described in the supplemental material to ref.~\cite{AguilarArevalo:2010wv}
    which we are following includes also MiniBooNE's sample of $\nu_\mu$
    events.  This is necessary to properly account for systematic uncertainties
    which are correlated between the two samples.  But of course, in a 3+1
    scenario, the $\nu_\mu$ sample suffers from $\nu_\mu$ disappearance into
    $\nu_s$, proportional to $|U_{\mu 4}|^2$. (Once again, in a 2-flavor model,
    only a much smaller fraction $\propto \sin^2 2\theta_{\mu e}$ will
    disappear, which is usually negligible.) The impact of including $\nu_\mu$
    disappearance is shown in panel (d) of \cref{fig:bg-osc}.
\end{enumerate}
We see that including the effect of 3+1 oscillations on the normalization
in the control regions and on the background prediction reduces the significance
of the MiniBooNE anomaly, though it remains above $3\sigma$.  These effects
are thus unable to fully explain the MiniBooNE anomaly, but they could well
be part of an ``Altarelli cocktail'' of several effects conspiring to lead to
the large observed excess~\cite{Brdar:inprep}.

Let us finally mention one caveat with the above corrections to the MiniBooNE
fit. Namely, we can only apply the corrections at the level of reconstructed
events as the mapping between true and reconstructed neutrino energies is not
publicly available for muon neutrinos. This means we have to assume that the
reconstructed neutrino energy is a faithful representation of the true neutrino
energy. While this is true for quasi-elastic scattering events which constitute
the majority of events, it is not the case for other event categories. For
instance, a neutrino--nucleon interaction may create an extra pion, and if this
pion is reabsorbed as it propagates out of the nucleus, the event will be
misinterpreted as a quasi-elastic interaction, and the kinematic reconstruction
of the neutrino energy based on the observed charged lepton energy and direction
will fail.

\section{Decay Widths and Transition Probability}
\label{sec:decay-widths}
 
Based on the interaction terms from \cref{eq:int-Lagrangian}, we can compute the
differential decay rates of the heavy neutrino $\nu_4$ and of the scalar $\phi$.
In the massless light neutrino limit, we obtain for the $\nu_4$ decay width
in the lab
\begin{align}
  \frac{1}{\tfrac{m_4}{E_4} \Gamma_{4}}
  \frac{d\Gamma^\text{lab}(\nu_4 \to \nu_j \phi)}{d E_j}
    &= \frac{|U_{sj}|^2}{\sum_{k=1}^3|U_{sk}|^2} \frac{E_j}{(1-x_{\phi 4}^2)^2 E_4^2} \,, \\
  \sum_j \frac{1}{\tfrac{m_4}{E_4} \Gamma_{4}}
  \frac{d\Gamma^\text{lab}(\nu_4 \to \nu_j \phi)}{d E_\phi}
    &= \frac{1}{1 - x_{\phi 4}^2} \frac{1}{E_4} \,.
  \label{eq:dGamma-nubar}
\end{align}
In these expressions, 
\begin{align}
  \Gamma_4 = \frac{g^2}{16 \pi} m_4 (1 - x_{\phi 4}^2)^2
             \sum_{j=1}^3 |U_{s4}^* U_{sj}|^2
\end{align}
is the total rest frame decay width of $\nu_4$, $x_{\phi 4} \equiv m_\phi / m_4$
is the ratio of
scalar and neutrino masses, and $E_j$, $E_\phi$ are the
daughter neutrino and scalar energies, respectively. In the $\nu_4$ rest frame,
$E_j$ is restricted to the interval $[0, m_4 (1-x_{\phi 4}^2)]$. 

The lab frame decay rate of the scalar $\phi$ is
\begin{align}
  \sum_{i,j} \frac{1}{\tfrac{m_\phi}{E_\phi} \Gamma_\phi}
    \frac{d\Gamma^\text{lab}(\phi \to \nu_i \bar{\nu}_j)}{d E_i}
     = \frac{1}{E_\phi} \,,
\end{align}
with the total rest frame decay width of $\phi$
\begin{align}
  \Gamma_\phi = \frac{g^2}{8 \pi} m_\phi \sum_{i, j=1}^3 |U_{s i}^* U_{sj}|^2 \,.
\end{align}
The kinematic constraint on the daughter neutrino energies is
$E_i, E_j \in [0, m_\phi]$.

\bibliographystyle{JHEP}
\bibliography{refs}

\providecommand{\href}[2]{#2}\begingroup\raggedright\begin{thebibliography}{10}

\bibitem{AguilarArevalo:2010wv}
{\bf MiniBooNE} Collaboration, A.~Aguilar-Arevalo et~al., {\it {Event Excess in
  the MiniBooNE Search for $\bar \nu_\mu \rightarrow \bar \nu_e$
  Oscillations}},  {\em Phys.Rev.Lett.} {\bf 105} (2010) 181801,
  [\href{http://arxiv.org/abs/1007.1150}{{\tt arXiv:1007.1150}}].

\bibitem{Aguilar-Arevalo:2013pmq}
{\bf MiniBooNE Collaboration} Collaboration, A.~Aguilar-Arevalo et~al., {\it
  {Improved Search for $\bar \nu_\mu \rightarrow \bar \nu_e$ Oscillations in
  the MiniBooNE Experiment}},  {\em Phys.Rev.Lett.} {\bf 110} (2013) 161801,
  [\href{http://arxiv.org/abs/1207.4809}{{\tt arXiv:1207.4809}}].

\bibitem{Aguilar-Arevalo:2018gpe}
{\bf MiniBooNE} Collaboration, A.~A. Aguilar-Arevalo et~al., {\it {Observation
  of a Significant Excess of Electron-Like Events in the MiniBooNE
  Short-Baseline Neutrino Experiment}},
  \href{http://arxiv.org/abs/1805.12028}{{\tt arXiv:1805.12028}}.

\bibitem{Aguilar:2001ty}
{\bf LSND} Collaboration, A.~Aguilar et~al., {\it {Evidence for neutrino
  oscillations from the observation of $\bar{\nu}_e$ appearance in a
  $\bar{\nu}_\mu$ beam}},  {\em Phys. Rev.} {\bf D64} (2001) 112007,
  [\href{http://arxiv.org/abs/hep-ex/0104049}{{\tt hep-ex/0104049}}].

\bibitem{Mention:2011rk}
G.~Mention, M.~Fechner, T.~Lasserre, T.~A. Mueller, D.~Lhuillier, M.~Cribier,
  and A.~Letourneau, {\it {The Reactor Antineutrino Anomaly}},  {\em Phys.Rev.}
  {\bf D83} (2011) 073006, [\href{http://arxiv.org/abs/1101.2755}{{\tt
  arXiv:1101.2755}}].

\bibitem{Dentler:2017tkw}
M.~Dentler, {\'A}.~Hern{\'a}ndez-Cabezudo, J.~Kopp, M.~Maltoni, and T.~Schwetz,
  {\it {Sterile Neutrinos or Flux Uncertainties? - Status of the Reactor
  Anti-Neutrino Anomaly}},  \href{http://arxiv.org/abs/1709.04294}{{\tt
  arXiv:1709.04294}}.

\bibitem{Acero:2007su}
M.~A. Acero, C.~Giunti, and M.~Laveder, {\it {Limits on nu(e) and anti-nu(e)
  disappearance from Gallium and reactor experiments}},  {\em Phys.Rev.} {\bf
  D78} (2008) 073009, [\href{http://arxiv.org/abs/0711.4222}{{\tt
  arXiv:0711.4222}}].

\bibitem{Giunti:2010zu}
C.~Giunti and M.~Laveder, {\it {Statistical Significance of the Gallium
  Anomaly}},  {\em Phys.Rev.} {\bf C83} (2011) 065504,
  [\href{http://arxiv.org/abs/1006.3244}{{\tt arXiv:1006.3244}}].

\bibitem{Kopp:2011qd}
J.~Kopp, M.~Maltoni, and T.~Schwetz, {\it {Are there sterile neutrinos at the
  eV scale?}},  {\em Phys.Rev.Lett.} {\bf 107} (2011) 091801,
  [\href{http://arxiv.org/abs/1103.4570}{{\tt arXiv:1103.4570}}].

\bibitem{Conrad:2012qt}
J.~Conrad, C.~Ignarra, G.~Karagiorgi, M.~Shaevitz, and J.~Spitz, {\it {Sterile
  Neutrino Fits to Short Baseline Neutrino Oscillation Measurements}},  {\em
  Adv.High Energy Phys.} {\bf 2013} (2013) 163897,
  [\href{http://arxiv.org/abs/1207.4765}{{\tt arXiv:1207.4765}}].

\bibitem{Archidiacono:2013xxa}
M.~Archidiacono, N.~Fornengo, C.~Giunti, S.~Hannestad, and A.~Melchiorri, {\it
  {Sterile Neutrinos: Cosmology vs Short-Baseline Experiments}},
  \href{http://arxiv.org/abs/1302.6720}{{\tt arXiv:1302.6720}}.

\bibitem{Kopp:2013vaa}
J.~Kopp, P.~A.~N. Machado, M.~Maltoni, and T.~Schwetz, {\it {Sterile Neutrino
  Oscillations: The Global Picture}},  {\em JHEP} {\bf 1305} (2013) 050,
  [\href{http://arxiv.org/abs/1303.3011}{{\tt arXiv:1303.3011}}].

\bibitem{Mirizzi:2013kva}
A.~Mirizzi, G.~Mangano, N.~Saviano, E.~Borriello, C.~Giunti, et~al., {\it {The
  strongest bounds on active-sterile neutrino mixing after Planck data}},
  \href{http://arxiv.org/abs/1303.5368}{{\tt arXiv:1303.5368}}.

\bibitem{Giunti:2013aea}
C.~Giunti, M.~Laveder, Y.~Li, and H.~Long, {\it {Pragmatic View of
  Short-Baseline Neutrino Oscillations}},  {\em Phys.Rev.} {\bf D88} (2013)
  073008, [\href{http://arxiv.org/abs/1308.5288}{{\tt arXiv:1308.5288}}].

\bibitem{Gariazzo:2013gua}
S.~Gariazzo, C.~Giunti, and M.~Laveder, {\it {Light Sterile Neutrinos in
  Cosmology and Short-Baseline Oscillation Experiments}},  {\em JHEP} {\bf
  1311} (2013) 211, [\href{http://arxiv.org/abs/1309.3192}{{\tt
  arXiv:1309.3192}}].

\bibitem{Collin:2016rao}
G.~H. Collin, C.~A. Arg{\"u}elles, J.~M. Conrad, and M.~H. Shaevitz, {\it
  {Sterile Neutrino Fits to Short Baseline Data}},
  \href{http://arxiv.org/abs/1602.00671}{{\tt arXiv:1602.00671}}.

\bibitem{Gariazzo:2017fdh}
S.~Gariazzo, C.~Giunti, M.~Laveder, and Y.~F. Li, {\it {Updated Global 3+1
  Analysis of Short-BaseLine Neutrino Oscillations}},
  \href{http://arxiv.org/abs/1703.00860}{{\tt arXiv:1703.00860}}.

\bibitem{Giunti:2017yid}
C.~Giunti, X.~P. Ji, M.~Laveder, Y.~F. Li, and B.~R. Littlejohn, {\it {Reactor
  Fuel Fraction Information on the Antineutrino Anomaly}},
  \href{http://arxiv.org/abs/1708.01133}{{\tt arXiv:1708.01133}}.

\bibitem{Dentler:2018sju}
M.~Dentler, {\'A}.~Hern{\'a}ndez-Cabezudo, J.~Kopp, P.~A.~N. Machado,
  M.~Maltoni, I.~Martinez-Soler, and T.~Schwetz, {\it {Updated global analysis
  of neutrino oscillations in the presence of eV-scale sterile neutrinos}},
  \href{http://arxiv.org/abs/1803.10661}{{\tt arXiv:1803.10661}}.

\bibitem{Liao:2018mbg}
J.~Liao, D.~Marfatia, and K.~Whisnant, {\it {MiniBooNE, MINOS+ and IceCube data
  imply a baroque neutrino sector}},
  \href{http://arxiv.org/abs/1810.01000}{{\tt arXiv:1810.01000}}.

\bibitem{Moulai:2019gpi}
M.~H. Moulai, C.~A. Arg{\"u}elles, G.~H. Collin, J.~M. Conrad, A.~Diaz, and
  M.~H. Shaevitz, {\it {Combining Sterile Neutrino Fits to Short Baseline Data
  with IceCube Data}},  \href{http://arxiv.org/abs/1910.13456}{{\tt
  arXiv:1910.13456}}.

\bibitem{PalomaresRuiz:2005vf}
S.~Palomares-Ruiz, S.~Pascoli, and T.~Schwetz, {\it {Explaining LSND by a
  decaying sterile neutrino}},  {\em JHEP} {\bf 0509} (2005) 048,
  [\href{http://arxiv.org/abs/hep-ph/0505216}{{\tt hep-ph/0505216}}].

\bibitem{Bai:2015ztj}
Y.~Bai, R.~Lu, S.~Lu, J.~Salvado, and B.~A. Stefanek, {\it {Three Twin
  Neutrinos: Evidence from LSND and MiniBooNE}},
  \href{http://arxiv.org/abs/1512.05357}{{\tt arXiv:1512.05357}}.

\bibitem{deGouvea:2019qre}
A.~de~Gouv\^ea, O.~L.~G. Peres, S.~Prakash, and G.~V. Stenico, {\it {On The
  Decaying-Sterile Neutrino Solution to the Electron (Anti)Neutrino Appearance
  Anomalies}},  \href{http://arxiv.org/abs/1911.01447}{{\tt arXiv:1911.01447}}.

\bibitem{Hannestad:2013ana}
S.~Hannestad, R.~S. Hansen, and T.~Tram, {\it {How secret interactions can
  reconcile sterile neutrinos with cosmology}},  {\em Phys.Rev.Lett.} {\bf 112}
  (2014) 031802, [\href{http://arxiv.org/abs/1310.5926}{{\tt
  arXiv:1310.5926}}].

\bibitem{Dasgupta:2013zpn}
B.~Dasgupta and J.~Kopp, {\it {A m{\'e}nage {\`a} trois of eV-scale sterile
  neutrinos, cosmology, and structure formation}},  {\em Phys.Rev.Lett.} {\bf
  112} (2014) 031803, [\href{http://arxiv.org/abs/1310.6337}{{\tt
  arXiv:1310.6337}}].

\bibitem{GonzalezGarcia:2005xw}
M.~C. Gonzalez-Garcia, F.~Halzen, and M.~Maltoni, {\it {Physics reach of
  high-energy and high-statistics icecube atmospheric neutrino data}},  {\em
  Phys. Rev.} {\bf D71} (2005) 093010,
  [\href{http://arxiv.org/abs/hep-ph/0502223}{{\tt hep-ph/0502223}}].

\bibitem{Moss:2017pur}
Z.~Moss, M.~H. Moulai, C.~A. Arg{\"u}elles, and J.~M. Conrad, {\it {Exploring a
  nonminimal sterile neutrino model involving decay at IceCube}},  {\em Phys.
  Rev.} {\bf D97} (2018), no.~5 055017,
  [\href{http://arxiv.org/abs/1711.05921}{{\tt arXiv:1711.05921}}].

\bibitem{GonzalezGarcia:2008ru}
M.~C. Gonzalez-Garcia and M.~Maltoni, {\it {Status of Oscillation plus Decay of
  Atmospheric and Long-Baseline Neutrinos}},  {\em Phys. Lett.} {\bf B663}
  (2008) 405--409, [\href{http://arxiv.org/abs/0802.3699}{{\tt
  arXiv:0802.3699}}].

\bibitem{Gninenko:2009ks}
S.~Gninenko, {\it {The MiniBooNE anomaly and heavy neutrino decay}},  {\em
  Phys.Rev.Lett.} {\bf 103} (2009) 241802,
  [\href{http://arxiv.org/abs/0902.3802}{{\tt arXiv:0902.3802}}].

\bibitem{Gninenko:2010pr}
S.~N. Gninenko, {\it {A resolution of puzzles from the LSND, KARMEN, and
  MiniBooNE experiments}},  {\em Phys.Rev.} {\bf D83} (2011) 015015,
  [\href{http://arxiv.org/abs/1009.5536}{{\tt arXiv:1009.5536}}].

\bibitem{Bertuzzo:2018itn}
E.~Bertuzzo, S.~Jana, P.~A.~N. Machado, and R.~{Zukanovich Funchal}, {\it {A
  Dark Neutrino Portal to Explain MiniBooNE}},
  \href{http://arxiv.org/abs/1807.09877}{{\tt arXiv:1807.09877}}.

\bibitem{Ballett:2018ynz}
P.~Ballett, S.~Pascoli, and M.~Ross-Lonergan, {\it {U(1)' mediated decays of
  heavy sterile neutrinos in MiniBooNE}},
  \href{http://arxiv.org/abs/1808.02915}{{\tt arXiv:1808.02915}}.

\bibitem{Jordan:2018qiy}
J.~R. Jordan, Y.~Kahn, G.~Krnjaic, M.~Moschella, and J.~Spitz, {\it {Severe
  Constraints on New Physics Explanations of the MiniBooNE Excess}},
  \href{http://arxiv.org/abs/1810.07185}{{\tt arXiv:1810.07185}}.

\bibitem{Fischer:2019fbw}
O.~Fischer, A.~Hernandez-Cabezudo, and T.~Schwetz, {\it {Explaining the
  MiniBooNE excess by a decaying sterile neutrino with mass in the 250 MeV
  range}},  \href{http://arxiv.org/abs/1909.09561}{{\tt arXiv:1909.09561}}.

\bibitem{Lindner:2001fx}
M.~Lindner, T.~Ohlsson, and W.~Winter, {\it {A Combined treatment of neutrino
  decay and neutrino oscillations}},  {\em Nucl. Phys.} {\bf B607} (2001)
  326--354, [\href{http://arxiv.org/abs/hep-ph/0103170}{{\tt hep-ph/0103170}}].

\bibitem{Huber:2004ka}
P.~Huber, M.~Lindner, and W.~Winter, {\it {Simulation of long-baseline neutrino
  oscillation experiments with GLoBES}},  {\em Comput. Phys. Commun.} {\bf 167}
  (2005) 195, [\href{http://arxiv.org/abs/hep-ph/0407333}{{\tt
  hep-ph/0407333}}].

\bibitem{Kopp:2006wp}
J.~Kopp, {\it {Efficient numerical diagonalization of Hermitian $3 \times 3$
  matrices}},  {\em Int. J. Mod. Phys.} {\bf C19} (2008) 523--548,
  [\href{http://arxiv.org/abs/physics/0610206}{{\tt physics/0610206}}]. Erratum
  ibid.\ {\bf C19} (2008) 845.

\bibitem{Huber:2007ji}
P.~Huber, J.~Kopp, M.~Lindner, M.~Rolinec, and W.~Winter, {\it {New features in
  the simulation of neutrino oscillation experiments with GLoBES 3.0}},  {\em
  Comput. Phys. Commun.} {\bf 177} (2007) 432--438,
  [\href{http://arxiv.org/abs/hep-ph/0701187}{{\tt hep-ph/0701187}}].

\bibitem{Armbruster:2002mp}
{\bf KARMEN} Collaboration, B.~Armbruster et~al., {\it {Upper limits for
  neutrino oscillations muon-anti-neutrino ---> electron-anti-neutrino from
  muon decay at rest}},  {\em Phys. Rev.} {\bf D65} (2002) 112001,
  [\href{http://arxiv.org/abs/hep-ex/0203021}{{\tt hep-ex/0203021}}].

\bibitem{Agafonova:2013xsk}
{\bf OPERA} Collaboration, N.~Agafonova et~al., {\it {Search for $\nu_\mu
  \rightarrow \nu_e$ oscillations with the OPERA experiment in the CNGS beam}},
   {\em JHEP} {\bf 07} (2013) 004, [\href{http://arxiv.org/abs/1303.3953}{{\tt
  arXiv:1303.3953}}]. [Addendum: JHEP07,085(2013)].

\bibitem{Antonello:2012fu}
M.~Antonello, B.~Baibussinov, P.~Benetti, E.~Calligarich, N.~Canci, et~al.,
  {\it {Experimental search for the LSND anomaly with the ICARUS LAr TPC
  detector in the CNGS beam}},  \href{http://arxiv.org/abs/1209.0122}{{\tt
  arXiv:1209.0122}}.

\bibitem{Farese:2014}
C.~Farese, {\it {Results from ICARUS}},  2014.
\newblock talk given at the Neutrino 2014 conference in Boston, slides
  available at
  \url{https://indico.fnal.gov/materialDisplay.py?contribId=265\&sessionId=18\&materialId=slides\&confId=8022}.

\bibitem{Antonello:2015jxa}
M.~Antonello, B.~Baibussinov, P.~Benetti, F.~Boffelli, A.~Bubak, et~al., {\it
  {Some conclusive considerations on the comparison of the ICARUS $\nu_\mu \to
  \nu_e$ oscillation search with the MiniBooNE low-energy event excess}},
  \href{http://arxiv.org/abs/1502.04833}{{\tt arXiv:1502.04833}}.

\bibitem{Borodovsky:1992pn}
L.~Borodovsky, C.~Chi, Y.~Ho, N.~Kondakis, W.-Y. Lee, et~al., {\it {Search for
  muon-neutrino oscillations $\nu_\mu \to \nu_e$ ($\bar\nu_\mu \to \bar\nu_e$)
  in a wide band neutrino beam}},  {\em Phys.Rev.Lett.} {\bf 68} (1992)
  274--277.

\bibitem{Diaz:2019fwt}
A.~Diaz, C.~A. Arg{\"u}elles, G.~H. Collin, J.~M. Conrad, and M.~H. Shaevitz,
  {\it {Where Are We With Light Sterile Neutrinos?}},
  \href{http://arxiv.org/abs/1906.00045}{{\tt arXiv:1906.00045}}.

\bibitem{Adamson:2017uda}
{\bf MINOS} Collaboration, P.~Adamson et~al., {\it {Search for sterile
  neutrinos in MINOS and MINOS+ using a two-detector fit}},  {\em Submitted to:
  Phys. Rev. Lett.} (2017) [\href{http://arxiv.org/abs/1710.06488}{{\tt
  arXiv:1710.06488}}].

\bibitem{Louis:2018yeg}
W.~C. Louis, {\it {Problems With the MINOS/MINOS+ Sterile Neutrino
  Muon-Neutrino Disappearance Result}},
  \href{http://arxiv.org/abs/1803.11488}{{\tt arXiv:1803.11488}}.

\bibitem{Bryman:2019ssi}
D.~A. Bryman and R.~Shrock, {\it {Improved Constraints on Sterile Neutrinos in
  the MeV to GeV Mass Range}},  \href{http://arxiv.org/abs/1904.06787}{{\tt
  arXiv:1904.06787}}.

\bibitem{Atre:2009rg}
A.~Atre, T.~Han, S.~Pascoli, and B.~Zhang, {\it {The Search for Heavy Majorana
  Neutrinos}},  {\em JHEP} {\bf 0905} (2009) 030,
  [\href{http://arxiv.org/abs/0901.3589}{{\tt arXiv:0901.3589}}].

\bibitem{Dragoun:2015oja}
O.~Dragoun and D.~V{\'e}nos, {\it {Constraints on the Active and Sterile
  Neutrino Masses from Beta-Ray Spectra: Past, Present and Future}},  {\em J.
  Phys.} {\bf 3} (2016) 77--113, [\href{http://arxiv.org/abs/1504.07496}{{\tt
  arXiv:1504.07496}}].

\bibitem{deGouvea:2015euy}
G.~de~Andr{\'e} and A.~Kobach, {\it {Global Constraints on a Heavy Neutrino}},
  \href{http://arxiv.org/abs/1511.00683}{{\tt arXiv:1511.00683}}.

\bibitem{Dolinski:2019nrj}
M.~J. Dolinski, A.~W.~P. Poon, and W.~Rodejohann, {\it {Neutrinoless
  Double-Beta Decay: Status and Prospects}},
  \href{http://arxiv.org/abs/1902.04097}{{\tt arXiv:1902.04097}}.

\bibitem{Cyburt:2015mya}
R.~H. Cyburt, B.~D. Fields, K.~A. Olive, and T.-H. Yeh, {\it {Big Bang
  Nucleosynthesis: 2015}},  {\em Rev. Mod. Phys.} {\bf 88} (2016) 015004,
  [\href{http://arxiv.org/abs/1505.01076}{{\tt arXiv:1505.01076}}].

\bibitem{Aghanim:2018eyx}
{\bf Planck} Collaboration, N.~Aghanim et~al., {\it {Planck 2018 results. VI.
  Cosmological parameters}},  \href{http://arxiv.org/abs/1807.06209}{{\tt
  arXiv:1807.06209}}.

\bibitem{Cyr-Racine:2013jua}
F.-Y. Cyr-Racine and K.~Sigurdson, {\it {Limits on Neutrino-Neutrino Scattering
  in the Early Universe}},  \href{http://arxiv.org/abs/1306.1536}{{\tt
  arXiv:1306.1536}}.

\bibitem{Forastieri:2017oma}
F.~Forastieri, M.~Lattanzi, G.~Mangano, A.~Mirizzi, P.~Natoli, and N.~Saviano,
  {\it {Cosmic microwave background constraints on secret interactions among
  sterile neutrinos}},  \href{http://arxiv.org/abs/1704.00626}{{\tt
  arXiv:1704.00626}}.

\bibitem{Forastieri:2019cuf}
F.~Forastieri, M.~Lattanzi, and P.~Natoli, {\it {Cosmological constraints on
  neutrino self-interactions with a light mediator}},
  \href{http://arxiv.org/abs/1904.07810}{{\tt arXiv:1904.07810}}.

\bibitem{Escudero:2019gfk}
M.~Escudero and M.~Fairbairn, {\it {Cosmological Constraints on Invisible
  Neutrino Decays Revisited}},  \href{http://arxiv.org/abs/1907.05425}{{\tt
  arXiv:1907.05425}}.

\bibitem{Escudero:2019gvw}
M.~Escudero and S.~J. Witte, {\it {A CMB Search for the Neutrino Mass Mechanism
  and its Relation to the $H_0$ Tension}},
  \href{http://arxiv.org/abs/1909.04044}{{\tt arXiv:1909.04044}}.

\bibitem{Lancaster:2017ksf}
L.~Lancaster, F.-Y. Cyr-Racine, L.~Knox, and Z.~Pan, {\it {A tale of two modes:
  Neutrino free-streaming in the early universe}},  {\em JCAP} {\bf 1707}
  (2017), no.~07 033, [\href{http://arxiv.org/abs/1704.06657}{{\tt
  arXiv:1704.06657}}].

\bibitem{Oldengott:2017fhy}
I.~M. Oldengott, T.~Tram, C.~Rampf, and Y.~Y.~Y. Wong, {\it {Interacting
  neutrinos in cosmology: exact description and constraints}},  {\em JCAP} {\bf
  1711} (2017), no.~11 027, [\href{http://arxiv.org/abs/1706.02123}{{\tt
  arXiv:1706.02123}}].

\bibitem{Song:2018zyl}
N.~Song, M.~C. Gonzalez-Garcia, and J.~Salvado, {\it {Cosmological constraints
  with self-interacting sterile neutrinos}},
  \href{http://arxiv.org/abs/1805.08218}{{\tt arXiv:1805.08218}}.

\bibitem{Kreisch:2019yzn}
C.~D. Kreisch, F.-Y. Cyr-Racine, and O.~Dor{\'e}, {\it {The Neutrino Puzzle:
  Anomalies, Interactions, and Cosmological Tensions}},
  \href{http://arxiv.org/abs/1902.00534}{{\tt arXiv:1902.00534}}.

\bibitem{Blinov:2019gcj}
N.~Blinov, K.~J. Kelly, G.~Z. Krnjaic, and S.~D. McDermott, {\it {Constraining
  the Self-Interacting Neutrino Interpretation of the Hubble Tension}},
  \href{http://arxiv.org/abs/1905.02727}{{\tt arXiv:1905.02727}}.

\bibitem{Esteban:2018azc}
I.~Esteban, M.~C. Gonzalez-Garcia, A.~Hernandez-Cabezudo, M.~Maltoni, and
  T.~Schwetz, {\it {Global analysis of three-flavour neutrino oscillations:
  synergies and tensions in the determination of $\theta_{23}$, $\delta_{CP}$,
  and the mass ordering}},  {\em JHEP} {\bf 01} (2019) 106,
  [\href{http://arxiv.org/abs/1811.05487}{{\tt arXiv:1811.05487}}]. NuFIT 4.1
  (2019), www.nu-fit.org.

\bibitem{Bertuzzo:2018ftf}
E.~Bertuzzo, S.~Jana, P.~A.~N. Machado, and R.~{Zukanovich Funchal}, {\it
  {Neutrino Masses and Mixings Dynamically Generated by a Light Dark Sector}},
  \href{http://arxiv.org/abs/1808.02500}{{\tt arXiv:1808.02500}}.

\bibitem{Zhao:2017wmo}
Y.~Zhao, {\it {Cosmology and time dependent parameters induced by a misaligned
  light scalar}},  {\em Phys. Rev.} {\bf D95} (2017), no.~11 115002,
  [\href{http://arxiv.org/abs/1701.02735}{{\tt arXiv:1701.02735}}].

\bibitem{Chu:2018gxk}
X.~Chu, B.~Dasgupta, M.~Dentler, J.~Kopp, and N.~Saviano, {\it {Sterile
  neutrinos with secret interactions---cosmological discord?}},  {\em JCAP}
  {\bf 1811} (2018), no.~11 049, [\href{http://arxiv.org/abs/1806.10629}{{\tt
  arXiv:1806.10629}}].

\bibitem{Farzan:2019yvo}
Y.~Farzan, {\it {Ultra-light scalar saving the 3+1 neutrino scheme from the
  cosmological bounds}},  \href{http://arxiv.org/abs/1907.04271}{{\tt
  arXiv:1907.04271}}.

\bibitem{Kolb:1987qy}
E.~W. Kolb and M.~S. Turner, {\it {Supernova SN 1987a and the Secret
  Interactions of Neutrinos}},  {\em Phys. Rev.} {\bf D36} (1987) 2895.

\bibitem{Berryman:2014qha}
J.~M. Berryman, A.~de~Gouvea, and D.~Hernandez, {\it {Solar Neutrinos and the
  Decaying Neutrino Hypothesis}},  {\em Phys. Rev.} {\bf D92} (2015), no.~7
  073003, [\href{http://arxiv.org/abs/1411.0308}{{\tt arXiv:1411.0308}}].

\bibitem{Maltoni:2003cu}
M.~Maltoni and T.~Schwetz, {\it {Testing the statistical compatibility of
  independent data sets}},  {\em Phys. Rev.} {\bf D68} (2003) 033020,
  [\href{http://arxiv.org/abs/hep-ph/0304176}{{\tt hep-ph/0304176}}].

\bibitem{Kostensalo:2019vmv}
J.~Kostensalo, J.~Suhonen, C.~Giunti, and P.~C. Srivastava, {\it {The gallium
  anomaly revisited}},  {\em Phys. Lett.} {\bf B795} (2019) 542--547,
  [\href{http://arxiv.org/abs/1906.10980}{{\tt arXiv:1906.10980}}].

\bibitem{An:2016srz}
{\bf Daya Bay} Collaboration, F.~P. An et~al., {\it {Improved Measurement of
  the Reactor Antineutrino Flux and Spectrum at Daya Bay}},
  \href{http://arxiv.org/abs/1607.05378}{{\tt arXiv:1607.05378}}.

\bibitem{Mueller:2011nm}
T.~Mueller, D.~Lhuillier, M.~Fallot, A.~Letourneau, S.~Cormon, et~al., {\it
  {Improved Predictions of Reactor Antineutrino Spectra}},  {\em Phys.Rev.}
  {\bf C83} (2011) 054615, [\href{http://arxiv.org/abs/1101.2663}{{\tt
  arXiv:1101.2663}}].

\bibitem{Huber:2011wv}
P.~Huber, {\it {On the determination of anti-neutrino spectra from nuclear
  reactors}},  {\em Phys.Rev.} {\bf C84} (2011) 024617,
  [\href{http://arxiv.org/abs/1106.0687}{{\tt arXiv:1106.0687}}].

\bibitem{Brdar:inprep}
V.~Brdar, J.~Kopp, and P.~Machado, {\it {An Altarelli Cocktail for MiniBooNE:
  can a conspiracy of Standard Model effects explain the MiniBooNE anomaly?}},
  . in preparation.

\end{thebibliography}\endgroup

\end{document}